\documentclass[12pt,preprint]{elsarticle} 
\usepackage{bm} 
\usepackage{amsmath,amssymb}
\usepackage{color}
\usepackage[english]{babel}
\usepackage{caption}
\usepackage{subcaption}
\usepackage{natbib}
\usepackage{tikz}
\usepackage{subfig}
\usetikzlibrary{arrows}
\usepackage{pstricks-add}
\usepackage{graphicx}
\usepackage{epstopdf}
\usepackage[toc,page]{appendix}
\biboptions{sort&compress,authoryear}

\begin{document}
\begin{frontmatter}

\title{Wave Propagation in Undulated Structural Lattices}

\author[DG]{G. Trainiti\corref{cor1}}

\author[DG]{J. J. Rimoli}

\author[DG,MS]{M. Ruzzene}

\address[DG]{Daniel Guggenheim School of Aerospace Engineering, Georgia Institute of Technology, 270 Ferst Dr, Atlanta, Georgia 30332, USA}
\address[MS]{George W. Woodruff School of Mechanical Engineering, Georgia Institute of Technology, 801 Ferst Dr, Atlanta, Georgia 30332, USA}
\cortext[cor1]{Corresponding author at: Daniel Guggenheim School of Aerospace Engineering, Georgia Institute of Technology, 270 Ferst Dr, Atlanta, Georgia 30332, USA. Email: gtrainiti@gatech.edu - Tel.:+1 404 786 8489}

\begin{abstract}
This work investigates wave propagation in undulated square structural lattices. The undulated pattern is obtained by imposing an initial curvature to the lattice's elements. The study considers both periodic undulated structures, in which the undulation is uniform throughout the structure as well as graded undulated patterns, in which the undulation is modulated within the lattice. Undulation is specifically considered in relation to its ability to induce anisotropy in the equivalent mechanical properties and to break the symmetry of the straight square lattice. Results show that wave motion is inhibited within specified frequency ranges owing to the generation of band gaps, and in specific directions as a result of the undulation-induced anisotropy.
\end{abstract}
\end{frontmatter}

\newpage
\section{Introduction}
Lattice structures are extensively used in many branches of engineering, such as civil, mechanical and aerospace. Of interest are lattices' mechanical vibration filtering \citep{Martinsson01022003} and wave steering capabilities \citep{Spadoni2009435}, which make them suitable for the design of mechanical metamaterials \citep{hussein2014dynamics,Deymier2013}. The lattice topologies that have been investigated include for example square, triangular, hexagonal, re-entrant, Kagome and chiral \citep{phani2006wave,Gonella2008125,Spadoni2009435,Casadei2013}, among others. Recently, different approaches have been proposed in order to achieve performance flexibility and properties tunability, for example multi-stable magneto-elastic lattice structures exploit the rearrangement of the topology due to the interaction between mechanical instabilities and magnetic forces, which leads to changes in wave propagation properties \citep{Schaeffer2015}. Another example are reconfigurable cell symmetry lattices, where relaxation of the unit cell symmetry is achieved by endowing the lattice with piezoelectric patches shunted to resonant circuits, then tuning the local stiffness of the structure by modifying the relative circuital characteristics \citep{Gonella2015}. Topology reconfiguration is an active topic of research in the field of soft materials \citep{BertoldiPRB2}, whereby mechanical and wave propagation properties adaptation is sought as a result of topological changes induced by externally applied loads~\citep{PhysRevB.88.014304,229201}. Similar patterns can be identified through a formal design process for structural lattice materials, whose properties differ substantially from equivalent straight beams with the same connectivity \citep{ADEM:ADEM201300064}. The presence of curved elements also radically affects the dynamic behavior of the structure, as previously shown for periodically undulated beams and plates \citep{Trainiti2015}.
 
This paper investigates the wave properties of undulated lattice structures by means of a numerical Bloch-Floquet analysis. Band diagrams and group velocity plots for periodic undulated lattices of different configurations, geometrical and material parameters are evaluated and discussed.
As expected, the results show that wave propagation properties of lattice structures are highly affected by the specific undulation pattern, which can be the result of a design process, but may also be triggered by static or dynamic loading. In contrast to straight lattices, where the bending stiffness of the elements is decoupled from and much smaller than the axial stiffness, in undulated structures longitudinal and flexural modes strongly interact to produce band gaps, wave speed reduction and wave directionality.

The paper is organized in five sections, including this introduction. In Section 2, a brief description of the relevant geometric parameters of the studied structures is given and of the unit cells identified for different configurations. Section 3 discusses the wave propagation analysis through Bloch Analysis and its numerical implementation. The results section (Section 4) investigates the performance of various configurations in terms of band gaps  and wave directionality. Finally, closing remarks are presented in Section 6.

\section{Theoretical background}
\subsection{Geometry of undulated lattices}
\label{sec:Geometry}
\begin{figure}
        \centering
        \begin{subfigure}[b]{0.90\textwidth}
                \includegraphics[width=\textwidth]{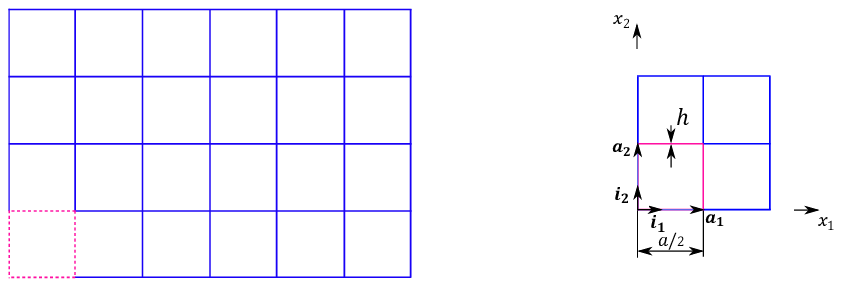}
                \caption{Straight configuration}
                \label{fig:UndulatedLatticeStriaght}
        \end{subfigure}      
        
        \begin{subfigure}[b]{0.90\textwidth}
                \includegraphics[width=\textwidth]{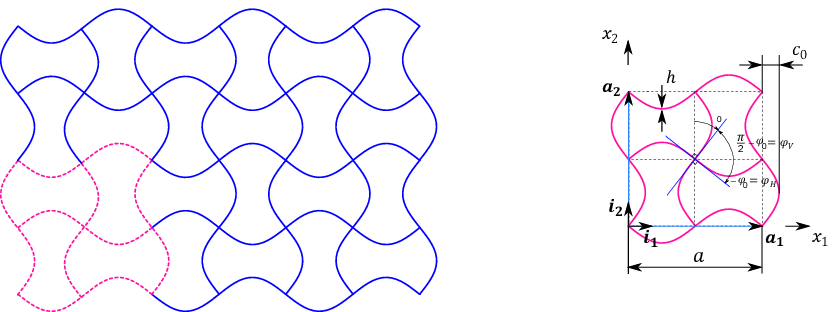}
                \caption{Undulated configuration 1}
                \label{fig:Conf1}
        \end{subfigure}                     

        \begin{subfigure}[b]{0.90\textwidth}
                \includegraphics[width=\textwidth]{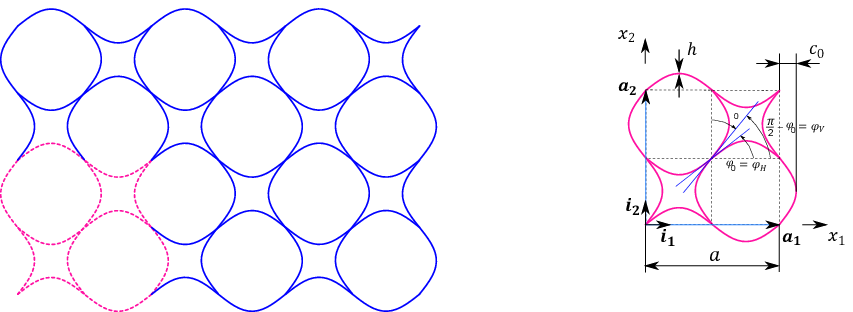}
                \caption{Undulated configuration 2}
                \label{fig:Conf2}
        \end{subfigure}                      
        \caption{Periodic straght and undulated lattice configurations 1 and 2 (left) with relative unit cells (right). The size of the undulated lattices unit cell is twice the size of the straight lattice unit cell.}\label{fig:UndulatedLatticeConfigurationsAndUC}
\end{figure}

The lattices considered are obtained by modifying a straight, square array of beam elements joined together at the intersections, as shown in Fig.~\ref{fig:UndulatedLatticeConfigurationsAndUC}. A Cartesian frame of reference is introduced with unit vectors $\hat{\mathbf{i}}_1$ and $\hat{\mathbf{i}}_2$, to define a set of lattices vectors $\mathbf{d}_1=a_1 \bm \hat{\mathbf{i}}_1, \mathbf{d}_2=a_2 \bm \hat{\mathbf{i}}_2$ that characterize periodic assemblies of unit cells (Fig.~\ref{fig:UndulatedLatticeConfigurationsAndUC}). Thus, the position of a generic lattice point $P$ is expressed as $\mathbf{r}_P=\mathbf{r}^l_{P} +n_1 \mathbf{d}_1+ n_2 \mathbf{d}_2$, which is defined by the local position of $P$ within the unit cell $\mathbf{r}^l_P=x^l_{1,P}\hat{\mathbf{i}}_1+x^l_{2,P}\hat{\mathbf{i}}_2$ and by a translation by integer multiples $n_1, n_2$ of the lattice vectors. For simplicity, a square lattice is considered, whereby $a_1=a_2=a$. Also, all beams have square cross section of thickness $h$.

\begin{figure}
  \centering
    \includegraphics[width=1\textwidth]{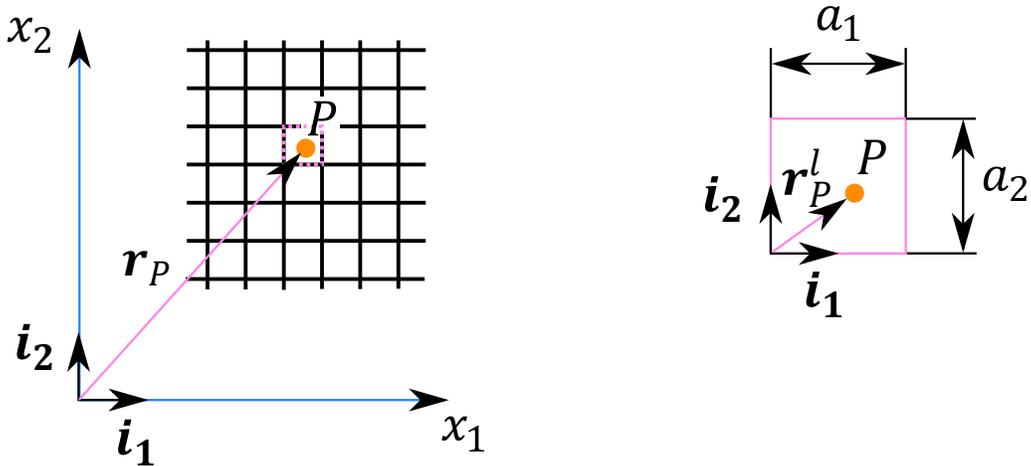}
      \caption{Position vector $\mathbf{r}_P$  and local position vector $\mathbf{r}^l_P$ of a point $P$ belonging to the lattice.}
      \label{fig:BlochAnalysis}
\end{figure}

Undulated lattice patterns are obtained by imposing an initial curvature to the beam elements. A harmonic undulation $c(x_i^l)=c \sin{\pi x_i^l /a}$ ($i=1,2$) is defined by the maximum distance $c$ between the centerlines of the undulated and the original straight configuration. The parameter $c$ can be held constant throughout the lattice, thus maintaining the periodicity of the structure, or can be modulated in space, i.e. letting $c=c(x_1,x_2)$, to produce spatially graded configurations.

The following dimensionless parameters are introduced:
\begin{equation}
\zeta=\frac{c}{a}
\label{exv:eqn:GeometricParameters1}
\end{equation}
where $\zeta$ defines the undulation amplitude relative to the unit cell size; and 
\begin{equation}
\gamma=\frac{h}{a}
\label{exv:eqn:GeometricParameters2}
\end{equation}
which is a slenderness parameter for the beams in the lattice.

Undulations imposed to an originally straight square lattice break geometrical symmetries. Two undulated configurations are obtained by considering two different angles at the corner node. Specifically, angles $\phi_H, \phi_V$ respectively identify lines tangent to the beams at the corner node: a first configuration (1) is characterized by $\phi_H=-\phi_0$, $\phi_V=\frac{\pi}{2}-\phi_0$, while the second configuration (2) corresponds to angles $\phi_H=\phi_0$, $\phi_V=\frac{\pi}{2}-\phi_0$, where $\phi_0 = \tan^{-1} (2\pi\zeta)$, as shown in Fig.~\ref{fig:UndulatedLatticeConfigurationsAndUC}. The ratio $r=\frac{\tan\phi_H}{\tan\phi_V}$, so that configuration 1 and configuration 2 are respectively characterized by $r_1=-\tan^2\phi_0=-(2\pi\zeta)^2$ and $r_2=\tan^2\phi_0=(2\pi\zeta)^2$, respectively, while $r=0$ for the straight lattice.

\section{Analysis of wave propagation}

\subsection{Bloch Analysis}
\begin{figure}
  \centering
    \includegraphics[width=0.5\textwidth]{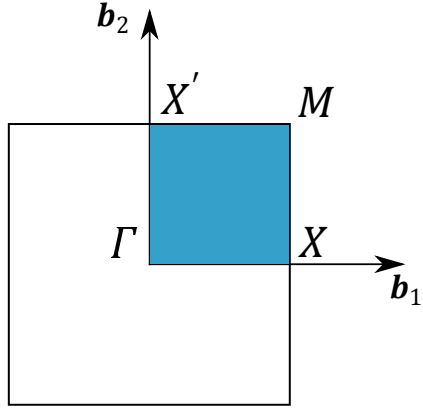}
      \caption{Basis vectors of the reciprocal lattice and First Brillouin Zone.}
      \label{fig:FBZ}
\end{figure}

The analysis of wave propagation is performed numerically by using the method suggested in \citep{aaberg1997usage}.
The displacement field $\mathbf{u}$ of point P of the lattice can be written in the form of a plane wave, leading to the following expression for the wave propagation solution: 
\begin{equation}
\mathbf{u}(\mathbf{r}_P,\omega) = \mathbf{u}_0 [\boldsymbol{\kappa}(\omega)] e^{j \boldsymbol{\kappa} \cdot \mathbf{r}_P}
\label{exv:Displacement}
\end{equation}
and
\begin{equation}
\mathbf{u}(\mathbf{r}_P,\omega) = \mathbf{u}^l [x^l_{1,P},x^l_{2,P},\boldsymbol{\kappa}(\omega)] e^{j \boldsymbol{\mu} \cdot \boldsymbol{n}}
\label{exv:BlochTheorem}
\end{equation}
where $\boldsymbol{\kappa}=\kappa_1\hat{\mathbf{i}}_1 + \kappa_2\hat{\mathbf{i}}_2$ is the wave vector, $\boldsymbol{\mu}=\kappa_1 a_1\hat{\mathbf{i}}_1 + \kappa_2 a_2\hat{\mathbf{i}}_2$ is the propagation vector while $\boldsymbol{n}=n_1\hat{\mathbf{i}}_1 + n_2 \hat{\mathbf{i}}_2$. The motion of a two-dimensional periodic structure is therefore expressed as the product of the response of a reference unit cell and a term giving information about the amplitude and the phase changes of the propagating wave when moving from one cell to the other. This result, expressed in Eq.~\ref{exv:BlochTheorem}, is known as Bloch's Theorem. Commonly, wave propagation properties of periodic structures are characterized by computing a relation between the periodicity of waves in space and time, namely the dispersion relation. In the case of a two-dimensional structure, the dispersion relation takes the form $\omega=\omega(\kappa_1,\kappa_2)$, a surface in the wavenumber domain, where the frequency $\omega$ is obtained as a function of the components of the wave vector $\boldsymbol{\kappa}$. Since these surfaces, called dispersion surfaces, are periodic in the wavenumber domain, it is possible to characterize the periodicity of the dispersion surfaces by defining a reciprocal lattice space and the corresponding basis vectors $\mathbf{b}_i$, which by definition satisfy $\mathbf{a}_i\cdot\mathbf{b}_j=2\pi\delta_{ij}$ for $i,j=1,2$ and $\delta_{ij}$ being the Kronecker delta. In the case of square direct lattice of spatial period $a$ in both directions $\hat{\mathbf{i}}_i$, the reciprocal lattice vectors are $\mathbf{b}_i=\frac{2 \pi}{a}\hat{\mathbf{i}}_i$ for $i=1,2$. Periodicity of the dispersion properties in the reciprocal lattice space is a key feature of periodic structure, for the calculation of the dispersion properties can be restricted to only selected areas of the reciprocal lattice space, depending on the goal of the analysis. If forbidden frequency ranges of the structure are investigated, i.e. band gaps, the dispersion relation is computed along the edges of the First Brillouin Zone (FBZ). In the case of a direct square lattice, it is clear that the corresponding reciprocal lattice will be square as well and the FBZ takes the form shown in Fig.~\ref{fig:FBZ}. Then it is customary to unfold the the resulting curve along the edges of the FBZ in order to obtain a planar representation, called band diagram.
When the directionality of wave propagation is of interest, the calculation of the dispersion relation is extended to the full FBZ. By taking the gradient of the dispersion surface one obtains the group velocity, $\boldsymbol{c}_g = c_{g_1}\hat{\mathbf{i}}_1 + c_{g_2}\hat{\mathbf{i}}_2 = \nabla \omega(\kappa_1,\kappa_2)$, which gives information about the direction of propagation of the energy throughout the structure by plotting the magnitude of $\boldsymbol{c}_g$ as a function of the angle in the usual Cartesian coordinate system of basis vectors $\hat{\mathbf{i}}_i$. Further details can be found in \citep{brillouin1953wave} and \citep{kittel2004introduction}.

\subsection{Evaluation of dispersion properties}
The unit cell is discretized using beam elements, whose behavior is described by 3 degrees of freedom: displacements 
along $\hat{\mathbf{i}}_1, \hat{\mathbf{i}}_2$ and a rotation about $\hat{\mathbf{i}}_3 = \hat{\mathbf{i}}_1 \times \hat{\mathbf{i}}_2$.
\begin{figure}
  \centering
    \includegraphics[width=0.5\textwidth]{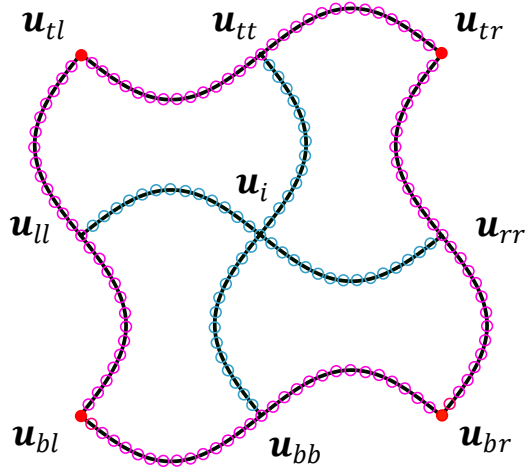}
      \caption{Discertized unit cell and node labeling.}
      \label{fig:DiscretizedUC}
\end{figure}
  
Applying Bloch's Theorem to the discretized unit cell corresponds to conveniently coupling the boundary nodes of the FE model. According to the node labeling in Fig.~\ref{fig:DiscretizedUC}, the set of equations to be enforced at the unit cell boundaries write:
\begin{align}
\label{eq:Bloch1}
\nonumber \mathbf{u}_{rr}  = \delta_1 \mathbf{u}_{ll}, \,\,
\mathbf{u}_{tt}  = \delta_2 \mathbf{u}_{bb}\\
\mathbf{u}_{bl} = \delta_1 \mathbf{u}_{bl}, \,\,
\mathbf{u}_{tl}  = \delta_2 \mathbf{u}_{vl} \\
\nonumber \mathbf{u}_{tr} = \delta_1\delta_2 \mathbf{u}_{bl}
\end{align}
where $\delta_i = e^{j \mu_{i}}$, $i=1,2$, with $\mu_{i} \in [0,\pi]$. Eigenfrequencies $\omega$ corresponding to the assigned dimensionless wavevector components $\mu_1$ and $\mu_2$ can be obtained by solving the following eigenvalue problem:
\begin{equation}
     \begin{bmatrix}
       \mathbf{Q}(\mu_1,\mu_2)
     \end{bmatrix}^T
\Bigg(
     \begin{bmatrix}
       \mathbf{K}& \mathbf{0}          \\[0.3em]
       \mathbf{0}& \mathbf{K}
     \end{bmatrix}
     -\omega^2
       \begin{bmatrix}
       \mathbf{M}& \mathbf{0}          \\[0.3em]
       \mathbf{0}& \mathbf{M}
     \end{bmatrix} 
\Bigg)
     \begin{bmatrix}
       \mathbf{Q}(\mu_1,\mu_2)
     \end{bmatrix}
     \begin{bmatrix}
       \mathbf{u}^{re}_i          \\[0.3em]
       \mathbf{u}^{re}_a          \\[0.3em]
        \mathbf{u}^{im}_i          \\[0.3em]
       \mathbf{u}^{im}_a          \\[0.3em]
     \end{bmatrix}    
     =
     \begin{bmatrix}
       \mathbf{0}
     \end{bmatrix}
     \label{eq:EVP}     
\end{equation}
where a doubled model having a 'real' and an 'imaginary' mesh is considered according to the procedure described in~\citep{aaberg1997usage}.  All analysis reported are performed by using the FE commercial code ABAQUS/STANDARD$^\circledR$. Each unit cell is discretized by 480 B21 beam elements. These elements allow for transverse shear deformation, which plays an important role for thick beam elements and when high frequency modes are considered. The sectional shear stiffness assumed in the analysis is $K=kGA$, with $G$ shear modulus, $A$ area of the cross section and $k=0.85$ shear factor for the rectangular cross section \citep{1966Cowper}.

Depending upon the goal of the analysis, the dispersion relation $\omega=\omega(\mu_1,\mu_2)$ can be computed imposing a wavevector spanning the entire First Brillouin Zone (FBZ) or its boundaries~\cite{hussein2014dynamics}.

\section{Results}

\subsection{Overview}

The following sections present results for periodic undulated lattices and for lattices with graded undulated configurations. The graded designs are obtained by varying the undulation parameter $c$ within the lattice area. This leads to a transition between different lattices which occurs gradually and may allow the combination of static and dynamic properties of different configurations within a single structural assemblies. All studies consider structures that are made of aluminum with Young's modulus $E=73$ GPa, Poisson's ratio $\nu=0.33$ and density $\rho=2700$ $\frac{kg}{m^3}$.

Frequencies values are presented in terms of the normalized frequency $\Omega$ defined as:
\begin{equation}
\label{eq:Omega}
\Omega = \frac{f a}{c_L}
\end{equation}
where $f$ is frequency, $a$ is the spatial period of the structure and $c_L=\sqrt{\frac{E}{\rho}}$ is the velocity of longitudinal waves. 

Parametric studies investigate the effect of the geometric parameter $\zeta$ and the slenderness parameter on frequency band gaps and wave directionality produced by anisotropy.

\subsection{Periodic undulated lattices}
\subsubsection{Band diagrams} \label{BandDiagrams}

\begin{figure}
        \centering
        \begin{subfigure}[b]{0.45\textwidth}
                \includegraphics[width=\textwidth]{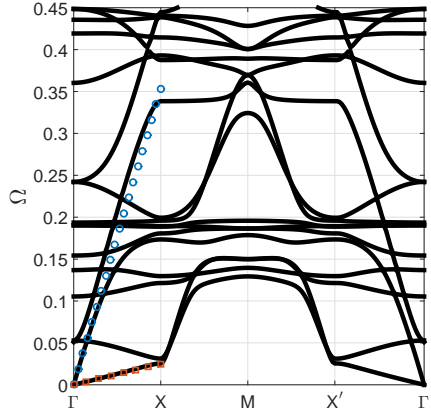}
                \caption{}
                \label{fig:BandDiagram_Straight_wLTdisp}
        \end{subfigure}\\
        \begin{subfigure}[b]{0.45\textwidth}
                \includegraphics[width=\textwidth]{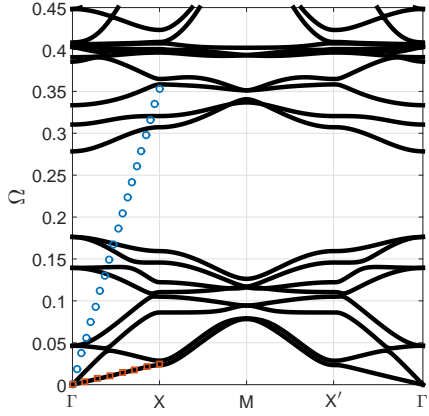}
                \caption{ }
                \label{fig:BandDiagram_Conf1_wLTdisp}
        \end{subfigure}
        \begin{subfigure}[b]{0.45\textwidth}
                \includegraphics[width=\textwidth]{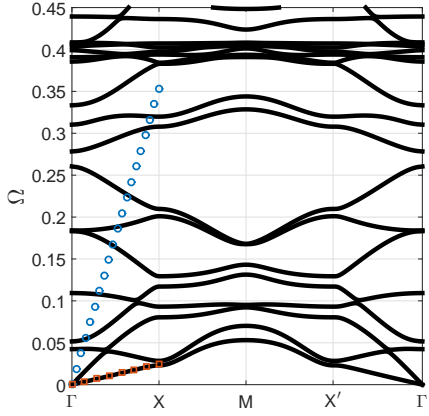}
                \caption{}
                \label{fig:BandDiagram_Conf2_wLTdisp}
        \end{subfigure}                       
        \caption{Band diagrams for straight lattice (Fig.~\ref{fig:BandDiagram_Straight_wLTdisp}), undulated configuration 1 (Fig.~\ref{fig:BandDiagram_Conf1_wLTdisp}), and undulated configuration 2 (Fig.~\ref{fig:BandDiagram_Conf2_wLTdisp})  ($\zeta=0.10$ and $\gamma=0.05$). Circular and square marker lines are the dispersion relations for the longitudinal and transverse modes in the square lattice expressed by eq.~(\ref{eq: square_velocities}).}\label{fig:BandDiagram_wLTdisp}
\end{figure}

Band diagrams of undulated lattices are compared with that of the straight square lattice with $\gamma=0.05$ shown in Fig.~\ref{fig:BandDiagram_Straight_wLTdisp}. The reduced symmetry induced by the undulations requires considering an enlarged 2x2 unit cell with respect to the straight lattice (Fig.~\ref{fig:UndulatedLatticeConfigurationsAndUC}). A first set of observations is made on the low frequency range, $\Omega \rightarrow 0$, \emph{i.e.} in the long wavelength limit. The two dispersion branches in the $\Gamma-X$ direction obey to the two analytical expressions:
\begin{equation}
\omega_l=c_l \kappa_l
\qquad
\omega_t=c_t \kappa_l
\label{eq:LongTransDispRel}
\end{equation}
and are associated with longitudinal and transverse waves in the lattice. In this frequency/wavelength range, the velocities $c_l$ and $c_t$ for a square lattice are given by \citep{phani2006wave,CellularSolids}:
\begin{equation}\label{eq: square_velocities}
c_l=\sqrt{\frac{E}{2 \rho}}
\qquad
c_t=\gamma\sqrt{\frac{E}{\rho}}
\end{equation}
where $E$ is the Young's modulus and $\rho$ is the density of the constituent material.

\begin{figure}
        \centering
        \begin{subfigure}[b]{0.65\textwidth}
                \includegraphics[width=\textwidth]{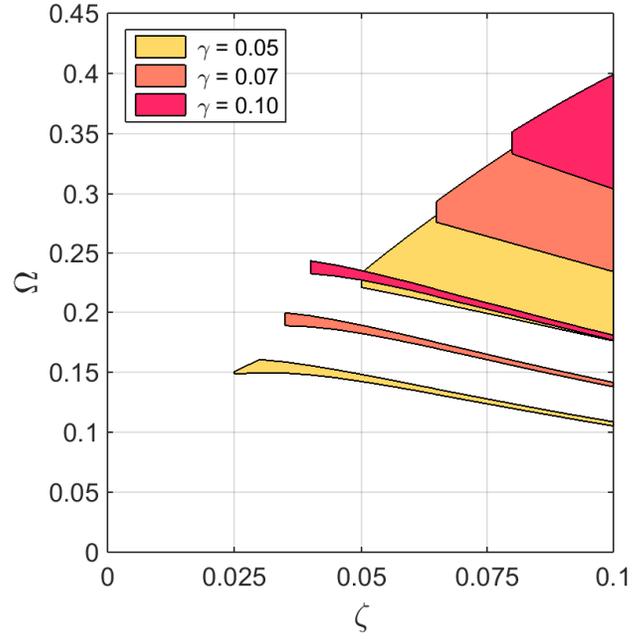}
                \caption{Configuration 1.}
                \label{fig:BandGapMap_Conf1}
        \end{subfigure}
        \quad \quad
        \begin{subfigure}[b]{0.65\textwidth}
                \includegraphics[width=\textwidth]{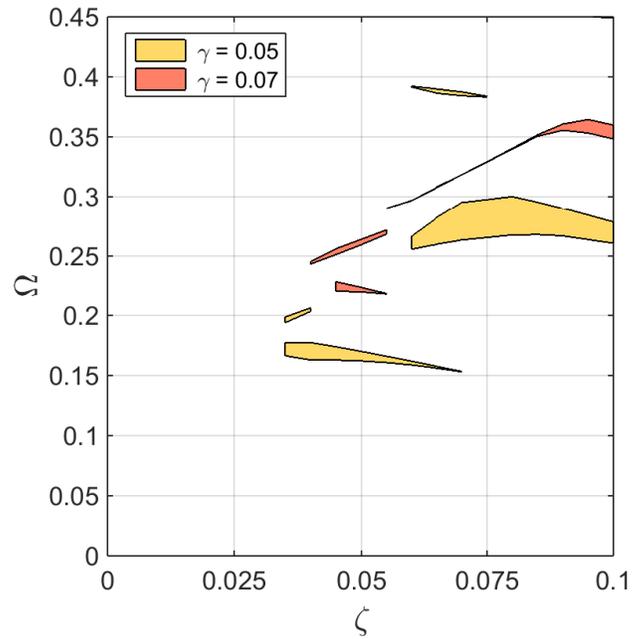}
                \caption{Configuration 2.}
                \label{fig:BandGapMap_Conf2}
        \end{subfigure}
                     
        \caption{Band gap maps of configurations 1 and 2 as a function of $\zeta$ and $\gamma$.}\label{fig:BandGapMaps}
\end{figure}

\begin{figure}
        \centering
        \begin{subfigure}[b]{0.18\textwidth}
                \includegraphics[width=\textwidth]{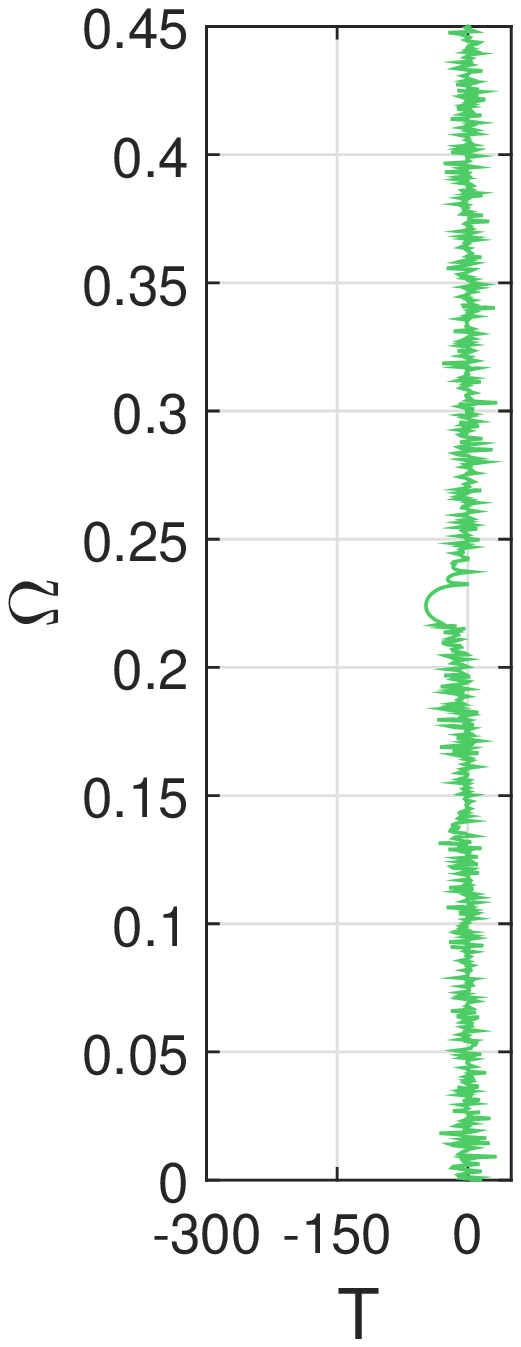}
                \caption{$\zeta=0.05$.}
                \label{fig:FR_Case2}
        \end{subfigure}
        \begin{subfigure}[b]{0.450\textwidth}
                \includegraphics[width=\textwidth]{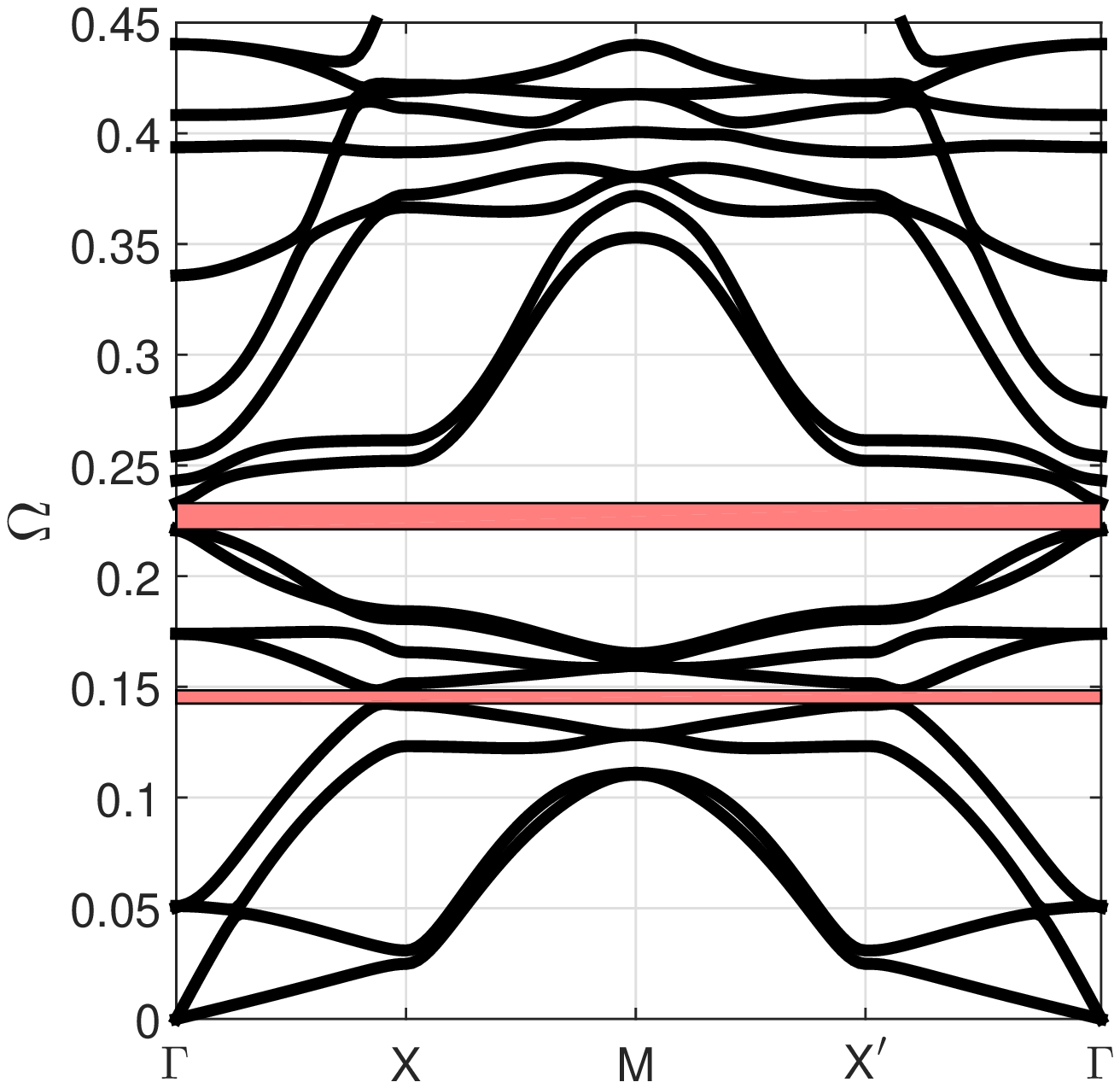}
                \caption{$\zeta=0.05$.}
                \label{fig:BandDiagramConf1_2}
        \end{subfigure}      

        \begin{subfigure}[b]{0.18\textwidth}
                \includegraphics[width=\textwidth]{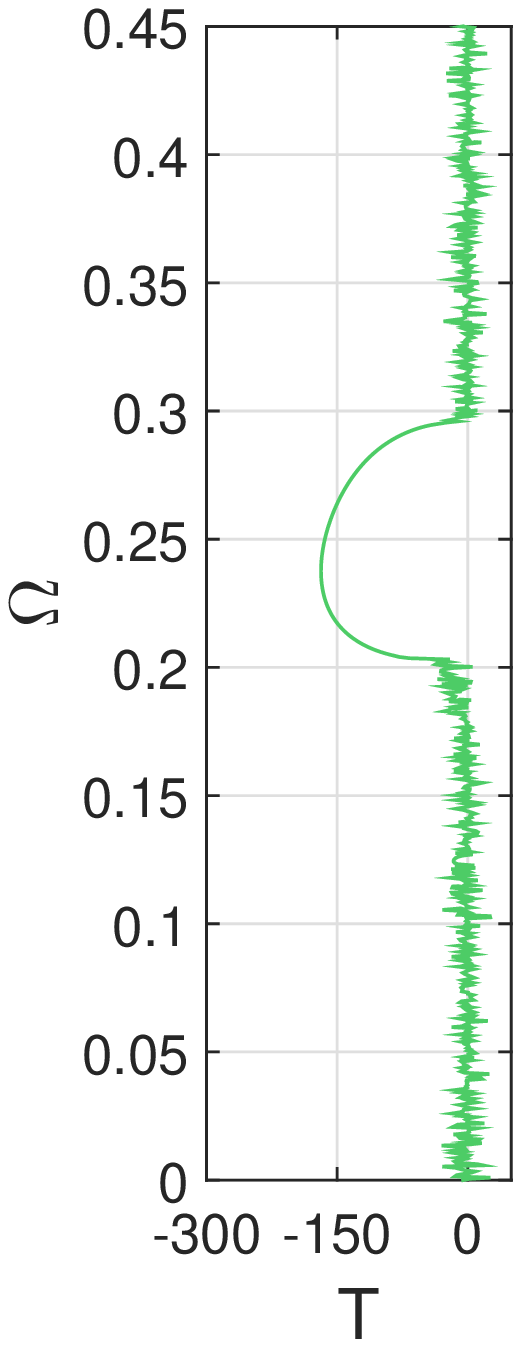}
                \caption{$\zeta=0.07$.}
                \label{fig:FR_Case4}
        \end{subfigure}
        \begin{subfigure}[b]{0.450\textwidth}
                \includegraphics[width=\textwidth]{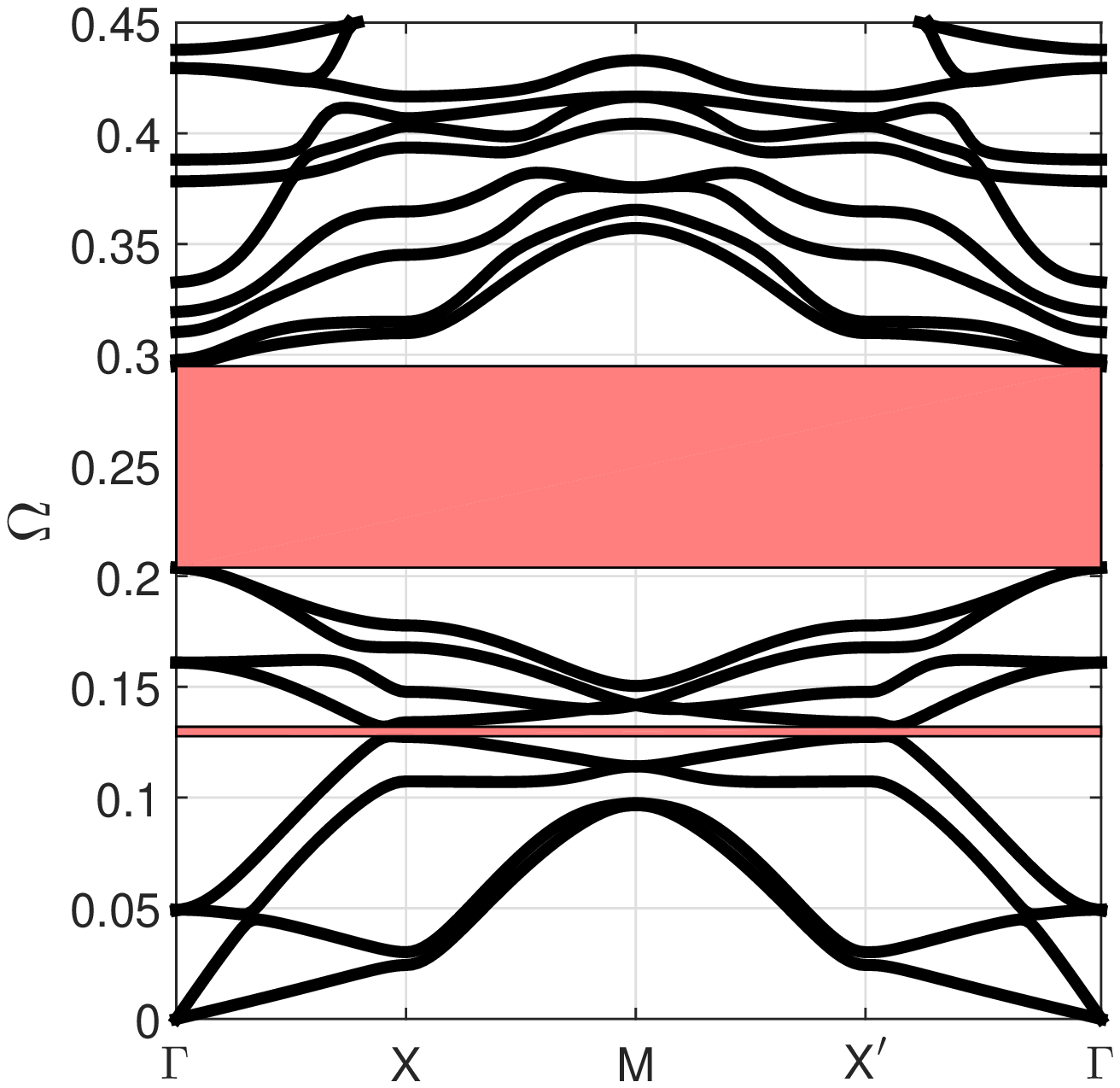}
                \caption{$\zeta=0.07$.}
                \label{fig:BandDiagramConf1_4}
        \end{subfigure}
        
        \begin{subfigure}[b]{0.18\textwidth}
                \includegraphics[width=\textwidth]{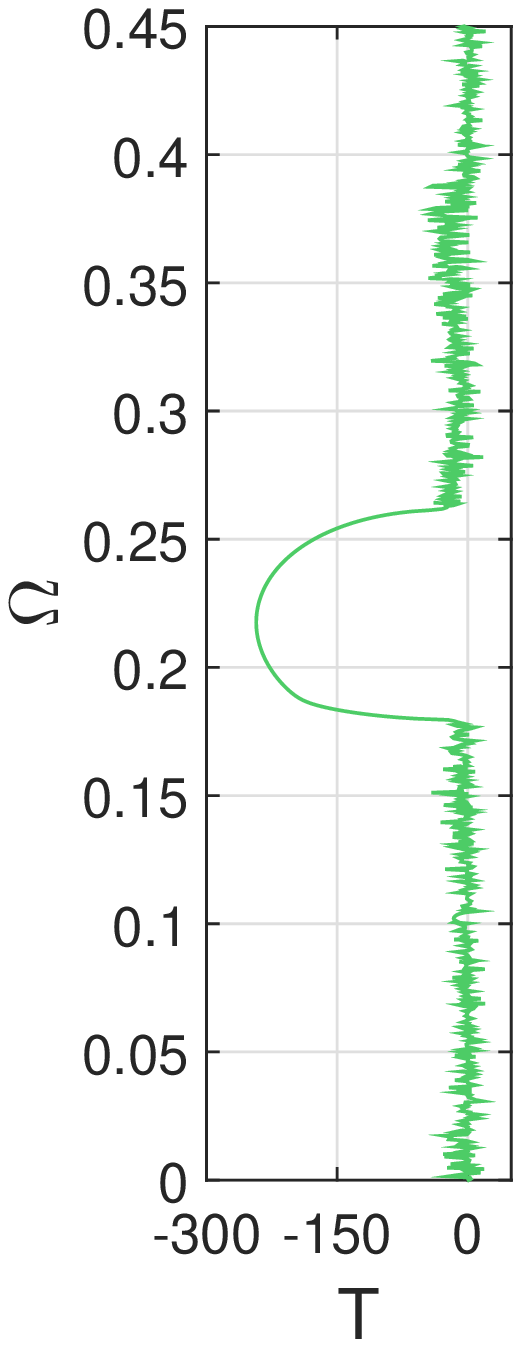}
                \caption{$\zeta=0.10$.}
                \label{fig:FR_Case6}
        \end{subfigure}
        \begin{subfigure}[b]{0.450\textwidth}
                \includegraphics[width=\textwidth]{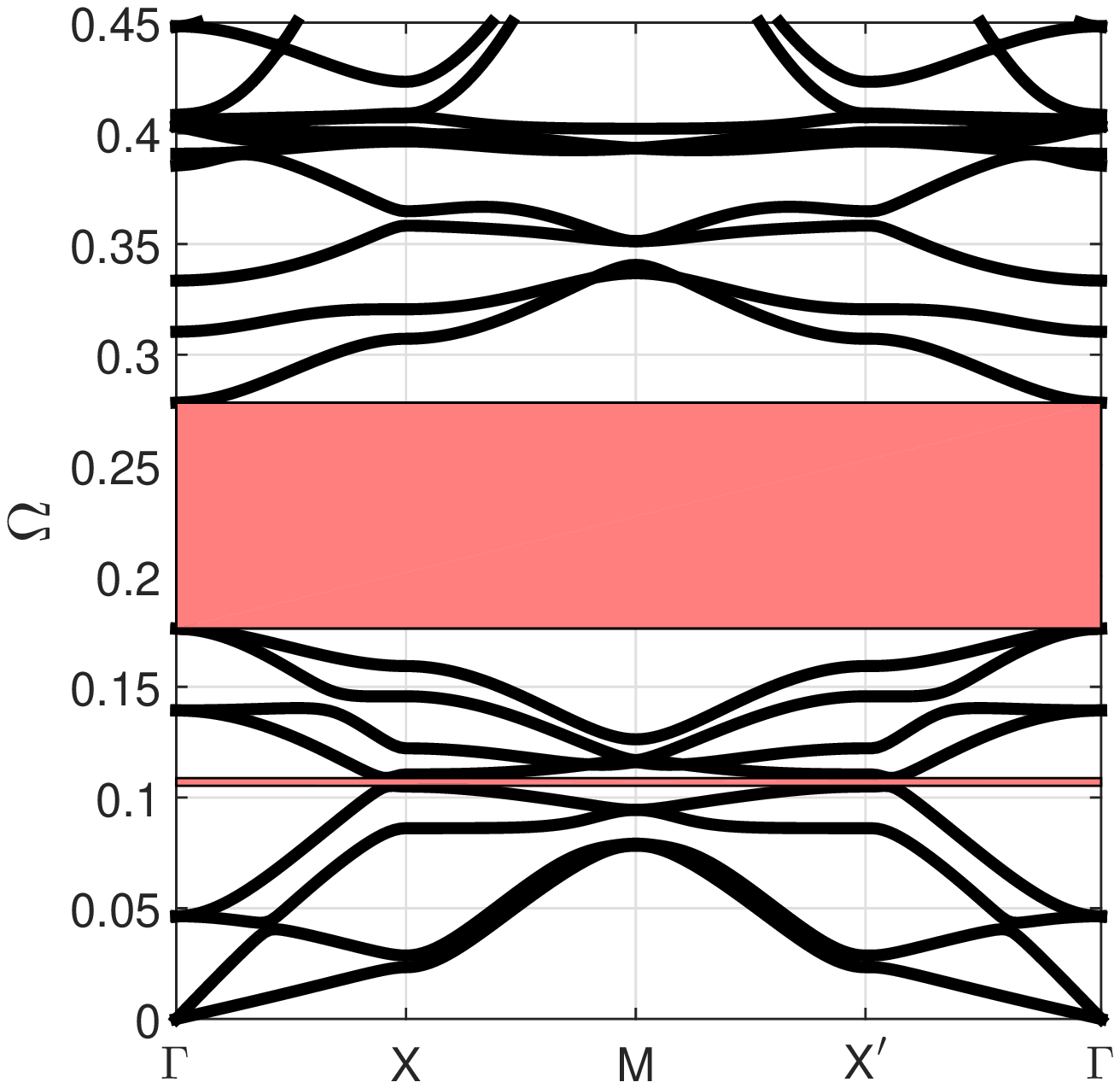}
                \caption{$\zeta=0.10$.}
                \label{fig:BandDiagramConf1_6}
        \end{subfigure}                     
        \caption{Frequency response diagrams (left) vs. band diagrams (right) of the undulated lattice in configuration 1 for $\gamma=0.05$ and increasing values of $\zeta$. The shaded areas correspond to band gaps.}\label{fig:FRvsBD_Conf1}
\end{figure}

\begin{figure}
        \centering
        \begin{subfigure}[b]{0.18\textwidth}
                \includegraphics[width=\textwidth]{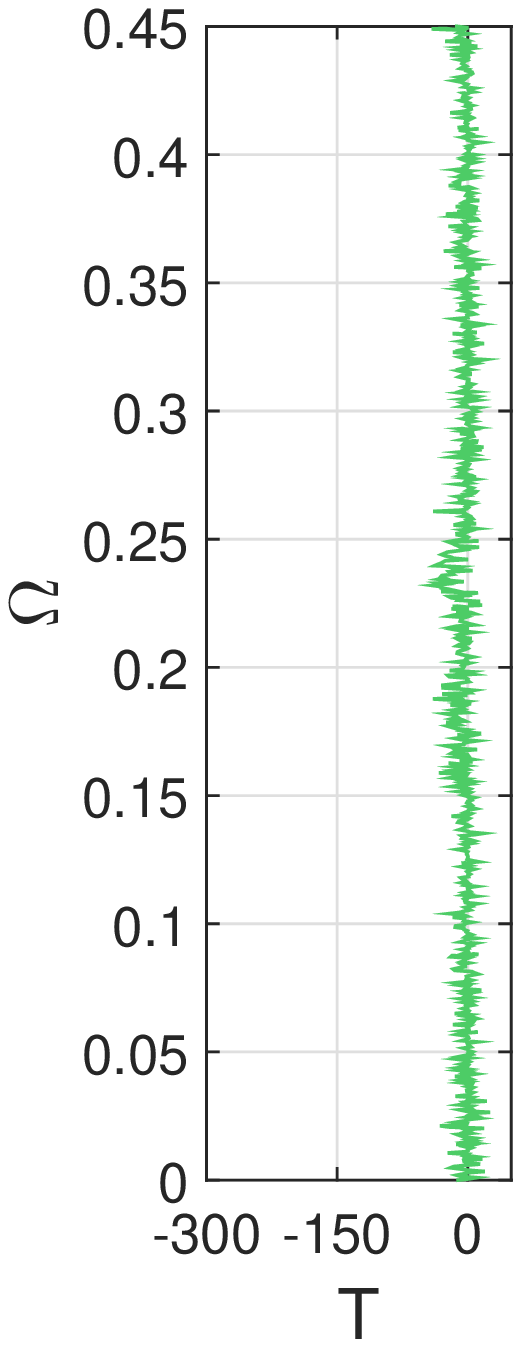}
                \caption{$\zeta=0.05$.}
                \label{fig:FR_Case3}
        \end{subfigure}
        \begin{subfigure}[b]{0.45\textwidth}
                \includegraphics[width=\textwidth]{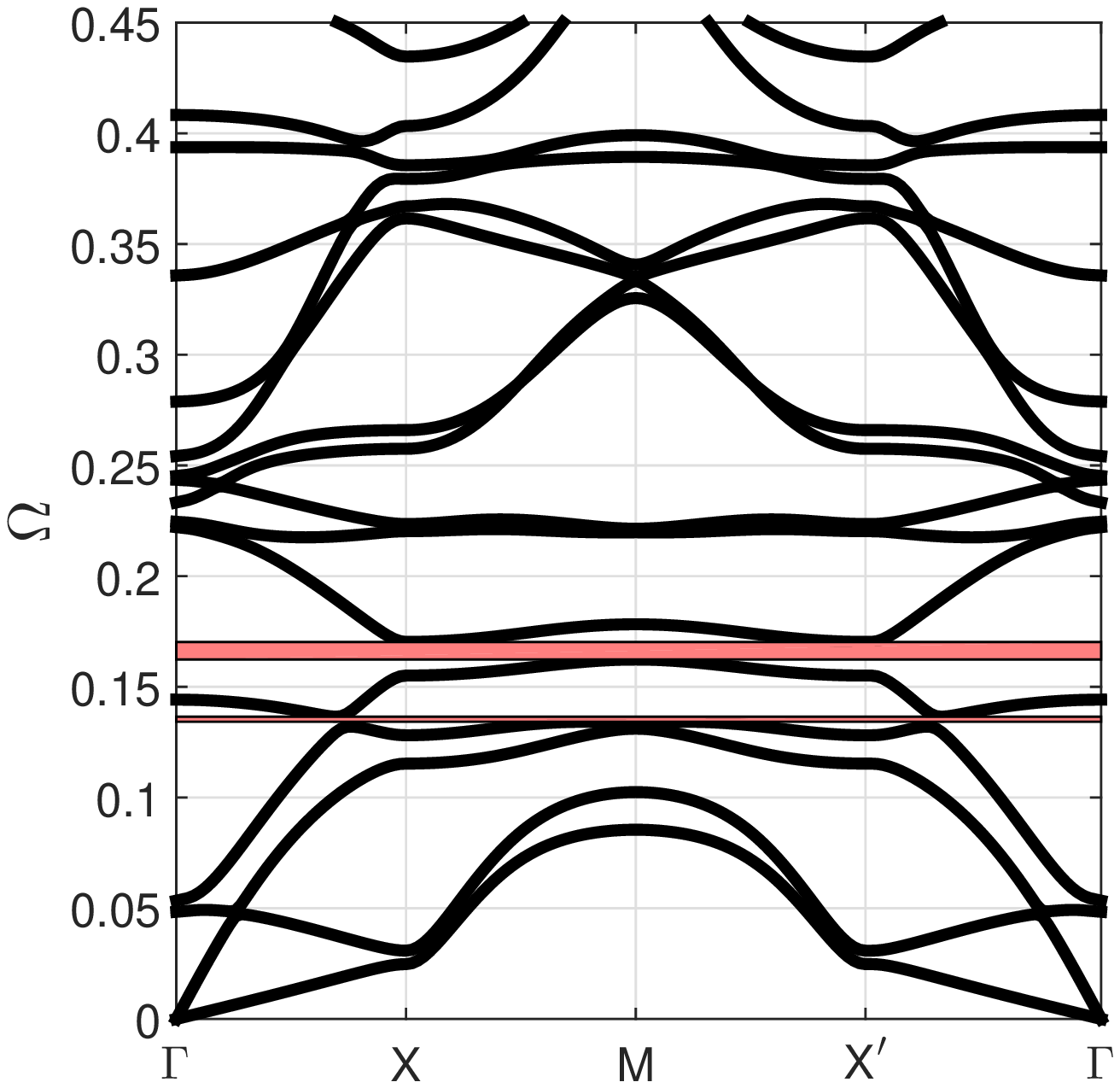}
                \caption{$\zeta=0.05$.}
                \label{fig:BandDiagramConf2_3}
        \end{subfigure}

        \begin{subfigure}[b]{0.18\textwidth}
                \includegraphics[width=\textwidth]{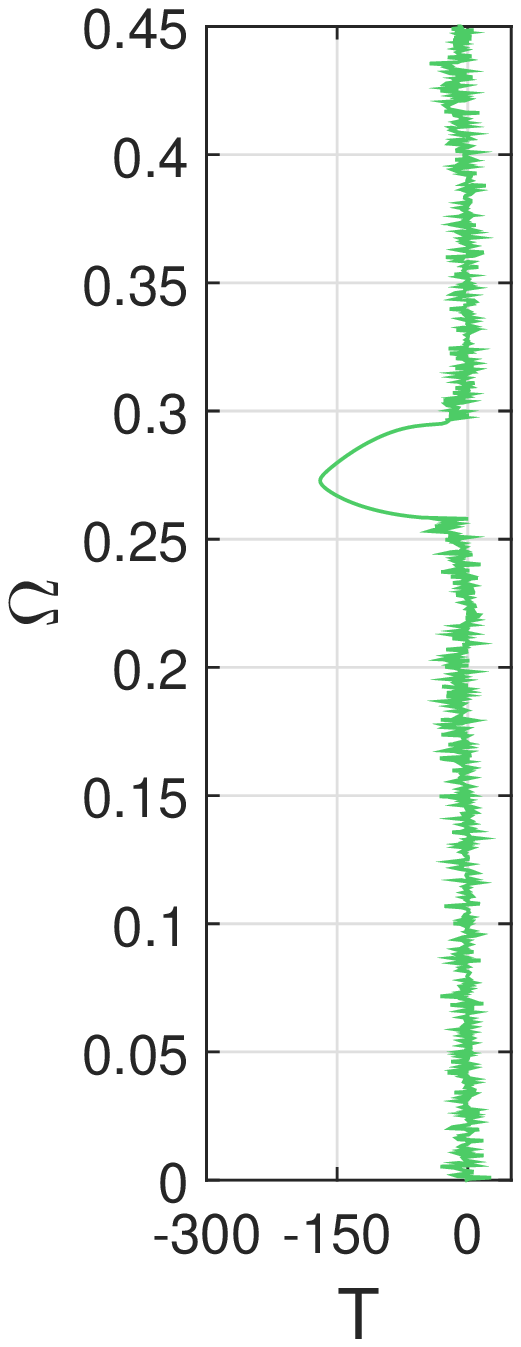}
                \caption{$\zeta=0.07$.}
                \label{fig:FR_Case5}
        \end{subfigure}
        \begin{subfigure}[b]{0.45\textwidth}
                \includegraphics[width=\textwidth]{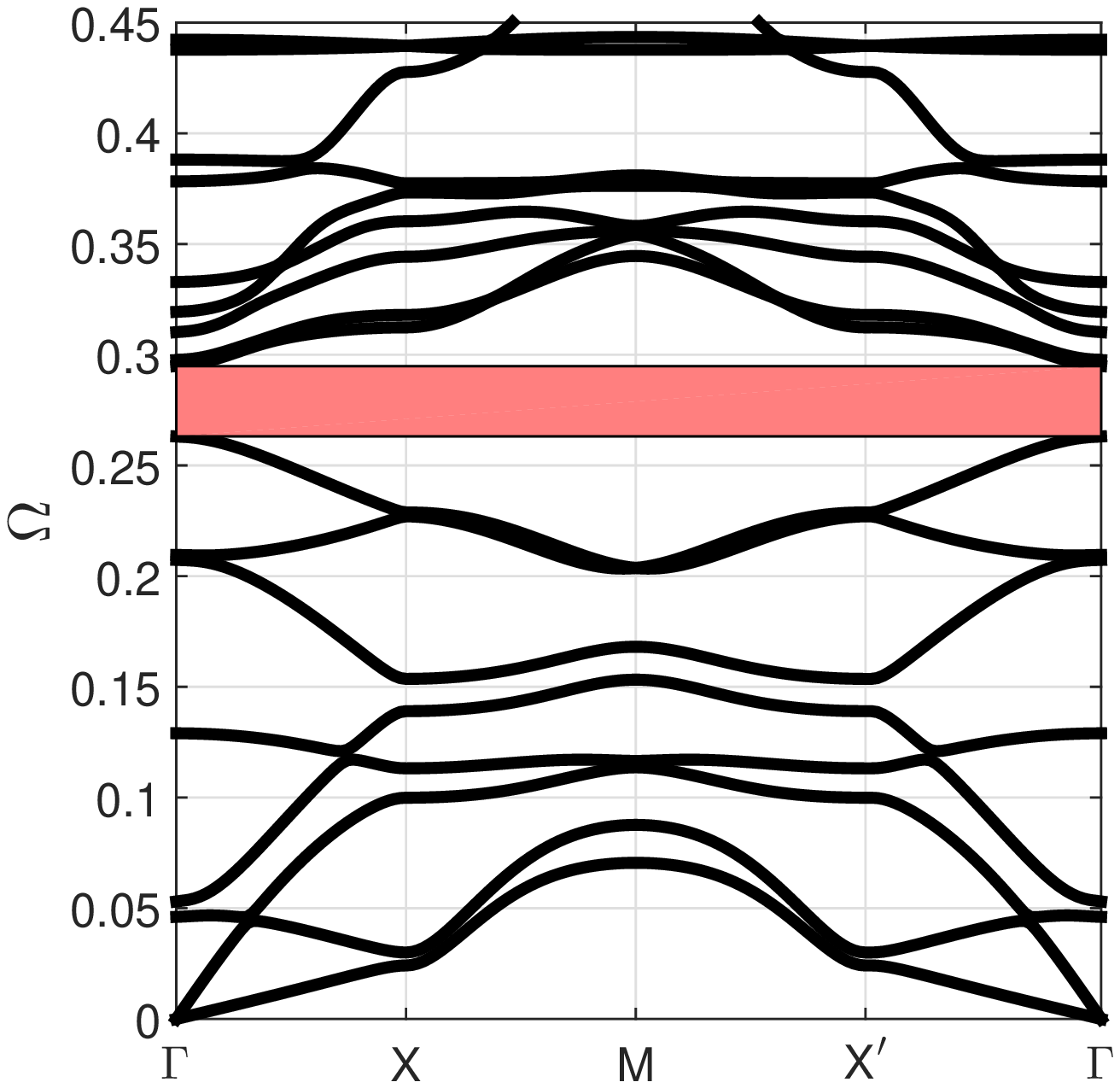}
                \caption{$\zeta=0.07$.}
                \label{fig:BandDiagramConf2_5}
        \end{subfigure}
        
        \begin{subfigure}[b]{0.18\textwidth}
                \includegraphics[width=\textwidth]{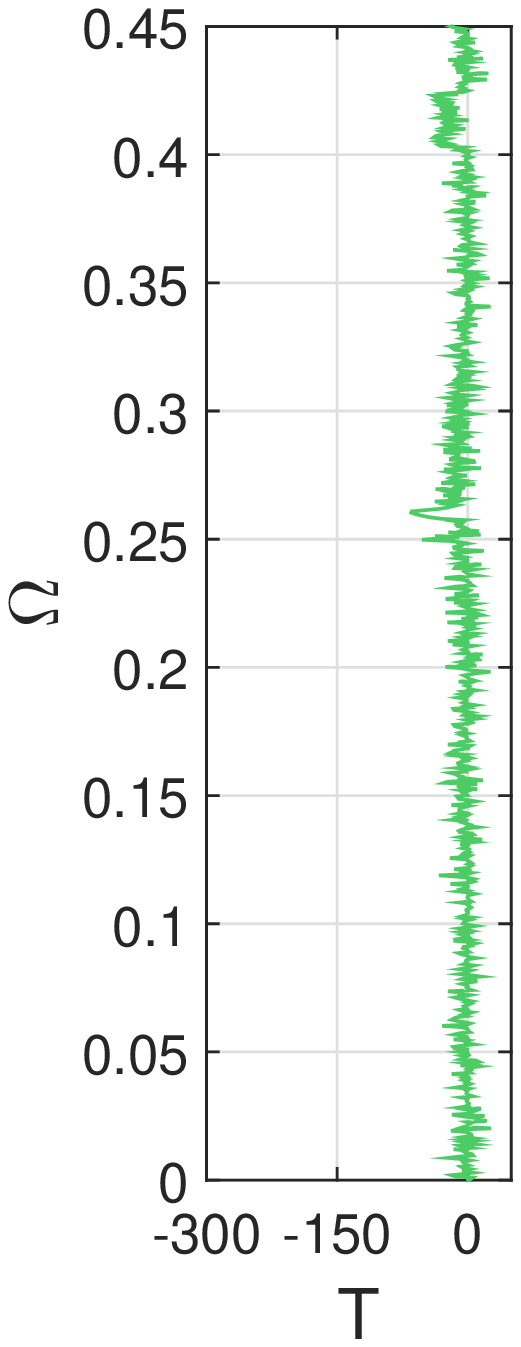}
                \caption{$\zeta=0.10$.}
                \label{fig:FR_Case7}
        \end{subfigure}
        \begin{subfigure}[b]{0.45\textwidth}
                \includegraphics[width=\textwidth]{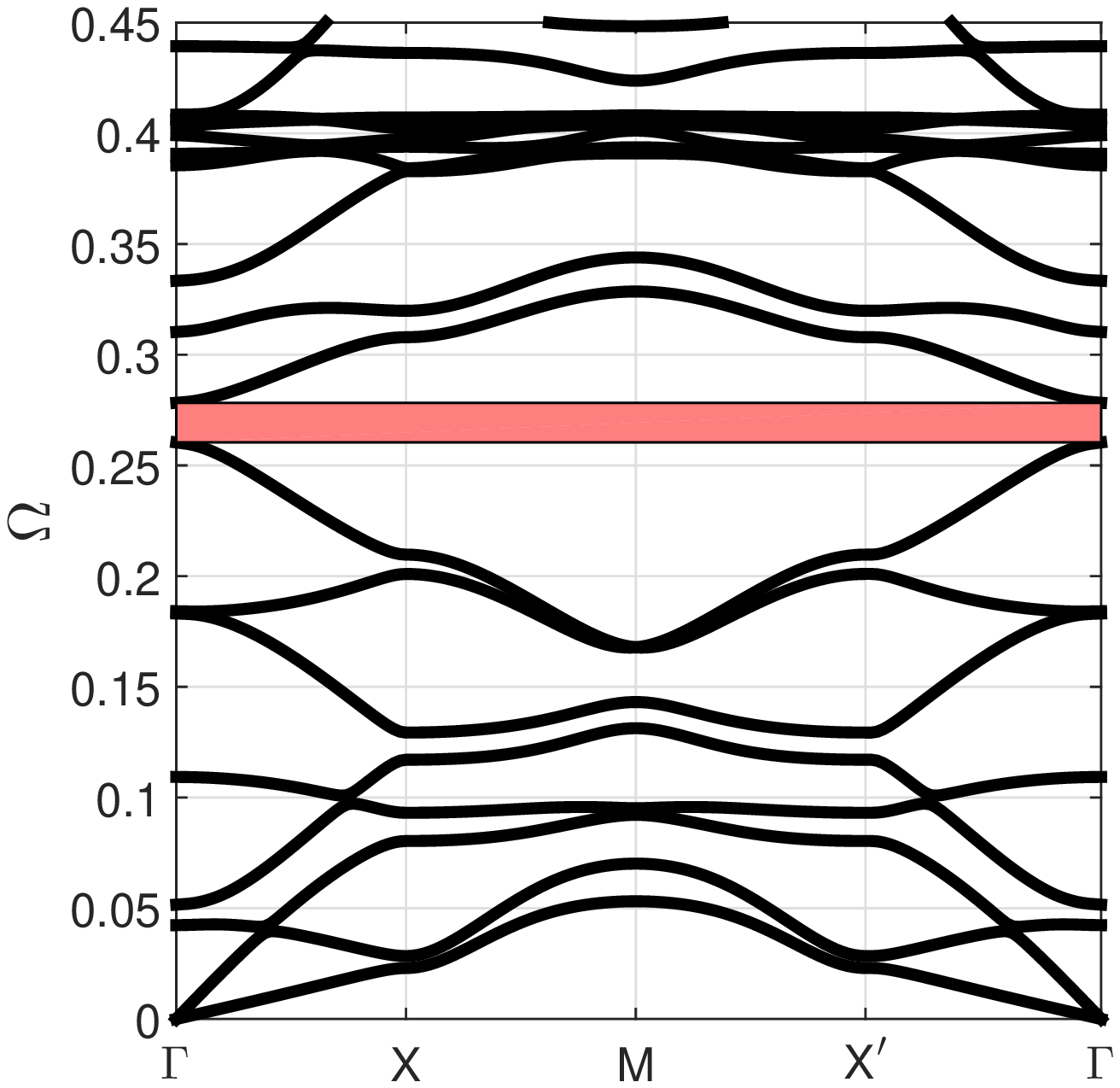}
                \caption{$\zeta=0.10$.}
                \label{fig:BandDiagramConf2_7}
        \end{subfigure}                     
        \caption{Frequency response diagrams (left) vs. band diagrams (right) of the undulated lattice in configuration 2 for $\gamma=0.05$ and increasing values of $\zeta$.}\label{fig:FRvsBD_Conf2}
\end{figure}

Striking differences between the straight and the undulated lattice configurations start to occur at higher frequencies. As already reported in \citep{phani2006wave}, straight lattices do not display band gaps in a wide range of slenderness parameter values $\gamma$. When undulation is introduced, large band gaps are obtained, as it can be seen in Fig.~\ref{fig:FRvsBD_Conf1} and Fig.~\ref{fig:FRvsBD_Conf2}. Although both undulated configurations considered display band gaps, the frequency at which they occur, their width and their behavior at different values of undulation amplitude are distinctive of the specific undulated configuration. Information about band gaps for different values of the geometric parameters are provided by the band gap maps in Fig.~\ref{fig:BandGapMaps}. These maps are obtained for both configurations by holding constant the values of $\gamma$ and $a$, and by varying the value of $\zeta$ within the range $[0,0.1]$. The undulated configuration 1 is characterized by two main band gaps, one that is less wide but opens at a lower value of the undulation amplitude $\zeta$ and a wider one that opens at a higher value of $\zeta$.
For increasing $\zeta$, the mean value of the wider band gap decreases, while its width first increases, then remains constant.
For undulated lattices of increasing slenderness parameter $\gamma$, hence for increasingly thicker beam elements, the two band gaps open at a higher value of undulation amplitude $\zeta$, also they are shifted at higher frequencies. On the other hand, band gaps for configuration 2 are less wide, and open and then close for increasing values of $\zeta$. Also, for thicker beam elements, they vanish almost completely in the range of parameters considered. In general, for a given configuration, each set of geometric parameters and each band gap defines a critical value $\zeta_c$ of the undulation amplitude corresponding to which a band gap opens as clearly shown in Fig.~\ref{fig:BandGapMaps}.

\begin{figure}
  \centering
    \includegraphics[width=0.75\textwidth]{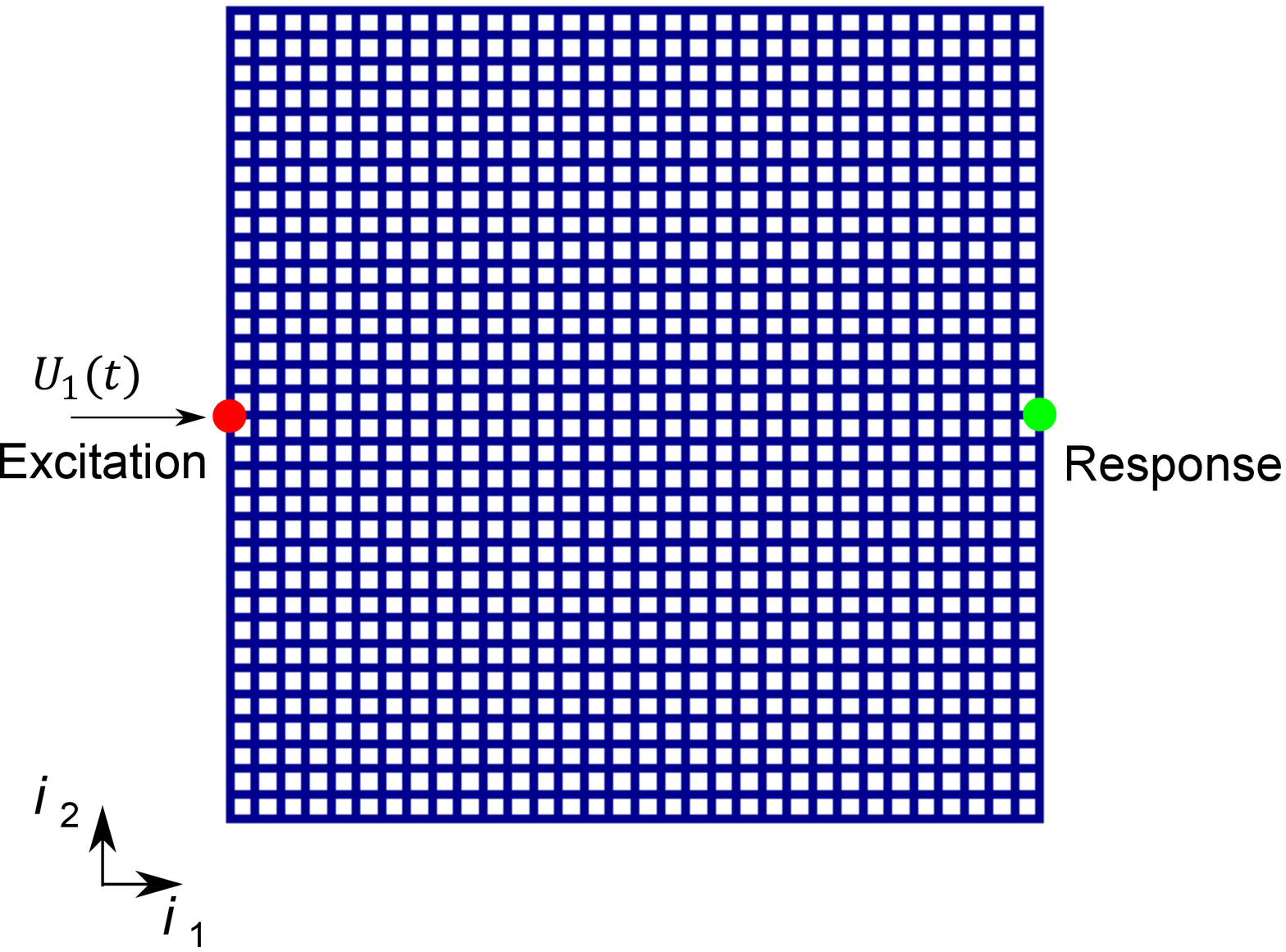}
      \caption{Schematics of a straight lattice subjected to imposed displacement $U_1(t)$ along the $\hat{\mathbf{i}}_1$ direction. The excitation is imposed at a node on the left side of the lattice's boundary, while the response is collected at a point located on the other side of the lattice.}
      \label{fig:SS_Schematics}
\end{figure}

A first verification of the analysis discussed so far is presented in the form of frequency response diagrams, shown on the left column of Fig.~\ref{fig:FRvsBD_Conf1} and \ref{fig:FRvsBD_Conf2}, which are obtained by calculating the steady-state dynamic response of finite 16x16 unit cell straight and undulated lattices. The responses are computed for harmonic excitation at a series of frequencies in the range  $\Omega \in [0,0.45]$. The excitation is enforced by imposing the displacement of a node of the boundary of the lattice along the $\hat{\mathbf{i}}_2$ direction, then collecting the magnitude of the response on the other side of the lattice, as shown in Fig.~\ref{fig:SS_Schematics}.

The plots display the transmission coefficient $T$ on the abscissa axis as a function of the non-dimensional frequency $\Omega$ on the ordinate axis. Transmission is computed as:
\[T=20Log_{10}(\frac{|U|}{|U|_{ref}})\]
where $|U|$ and $|U|_{ref}$ are the magnitude of the displacement vector at the point where the response is collected and at the point where the excitation is imposed, respectively. The coefficient $T$ provides information on the frequencies at which waves are allowed to propagate throughout the structure: low values of $T$ correspond to frequencies of forbidden wave propagation, since the perturbation applied locally remains local and decays quickly in the proximity of its point of application. Consistently with the information given by the band diagram in Fig.~\ref{fig:BandDiagram_Straight_wLTdisp}, the values of the parameter $T$ show that the straight lattice with $\gamma=0.05$ allows wave propagation in all the considered frequency range.
Conversely, in undulated lattices a significant drop of T is associated with frequencies corresponding to the band gaps represented by shaded areas in the diagrams of Fig.~\ref{fig:FRvsBD_Conf1} and Fig.~\ref{fig:FRvsBD_Conf2}.

\subsubsection{Directionality of wave propagation: group velocities and anisotropy index}
The second step of the analysis characterizes the directionality of wave motion, which is predicted from the evaluation of the group velocity directional dependence. A quantitative measure is introduced in the form of a dimensionless index named the anisotropy index ($AI$), as first introduced in~\citep{Casadei2013}. Transient responses are computed to illustrate the manifestation of directionality in finite lattices. The two undulated lattice configurations with $\gamma=0.05$ are investigated, both being characterized by $\zeta=0.1$. 

It is known that square honeycombs, contrary to other lattice structures such as triangular, hexagonal honeycombs or Kagome lattices, are strongly anisotropic at low frequencies, with transverse modes (S-modes) displaying a larger degree of anisotropy than longitudinal modes (L-modes)~\citep{phani2006wave,Casadei2013}. As a result of this anisotropy, the elastic energy propagates throughout the structure with different velocities depending on the direction of propagation. At higher frequencies, combinations of S-modes and L-modes generate complex wave propagation patterns, where multiple modes are characterized by different degrees of anisotropy. Information on wave propagation anisotropy of a structure can be obtained by plotting the components of the group velocity $\boldsymbol{c}_g$ associated to a certain value of frequency, which reveals the difference in velocity of propagation in different directions. When the components of $\boldsymbol{c}_g$ corresponding to a certain wave mode lead to a circular plot, no preferential direction for the energy is observed,  and the structure is isotropic. In contrast, polar group velocity plots departing from circularity characterize non-isotropic behavior.

A quantitative characterization of anisotropy relies on the evaluation of the angular dependence of the magnitude $c_g = |\boldsymbol{c}_g|$ of the group velocity in relation to its mean circumferential value. This leads to the formulation of the anisotropy index ($AI$), which is defined as~\citep{Casadei2013}:

\begin{equation}
AI=\sqrt{\int_{0}^{2 \pi} \Big[\frac{c_g(\theta) - \bar{c}_g}{\bar{c}_g}\Big]^2 d\theta}
\label{eq:AI}
\end{equation}
where $c_g(\theta)$ denotes the magnitude of the group velocity relative to a certain mode as a function of the angular position $\theta$, while $\overline{c}_g $ is the average value of $c_g(\theta)$ over the full $360^\circ$ angular range, i.e.:
\begin{equation}
\overline{c}_g = \frac{1}{2 \pi} \int_{0}^{2 \pi} c_g (\theta) d\theta
\label{eq:cgbar}
\end{equation}

\begin{figure}
        \centering
        \begin{subfigure}[b]{0.5\textwidth}
                \includegraphics[width=\textwidth]{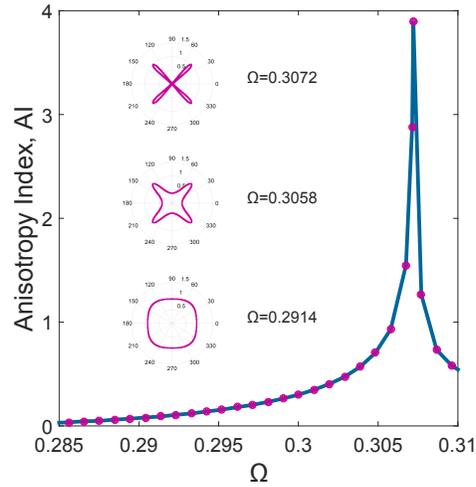}
                \caption{Anisotropy index.}
                \label{fig:AI_Conf1_Polar15150Hz15900Hz15974Hz}
        \end{subfigure}
        \quad \quad
        \begin{subfigure}[b]{0.5\textwidth}
                \includegraphics[width=\textwidth]{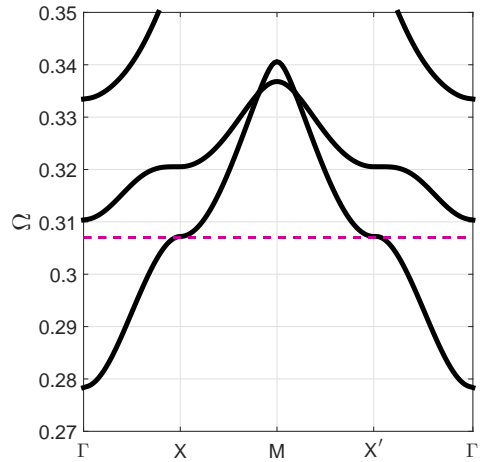}
                \caption{Band Diagram.}
                \label{fig:BD_UndConf1_Directionality}
        \end{subfigure}
                     
        \caption{Anisotropy index, polar group velocity plots for Configurations 1 (left) and detail of the band diagram (right). Geometric parameters: $\gamma=0.05$ and $\zeta=0.1$. Polar plots are normalized.}\label{fig:AI_BD_Conf1}
\end{figure}

\begin{figure}
        \centering
        \begin{subfigure}[b]{0.5\textwidth}
                \includegraphics[width=\textwidth]{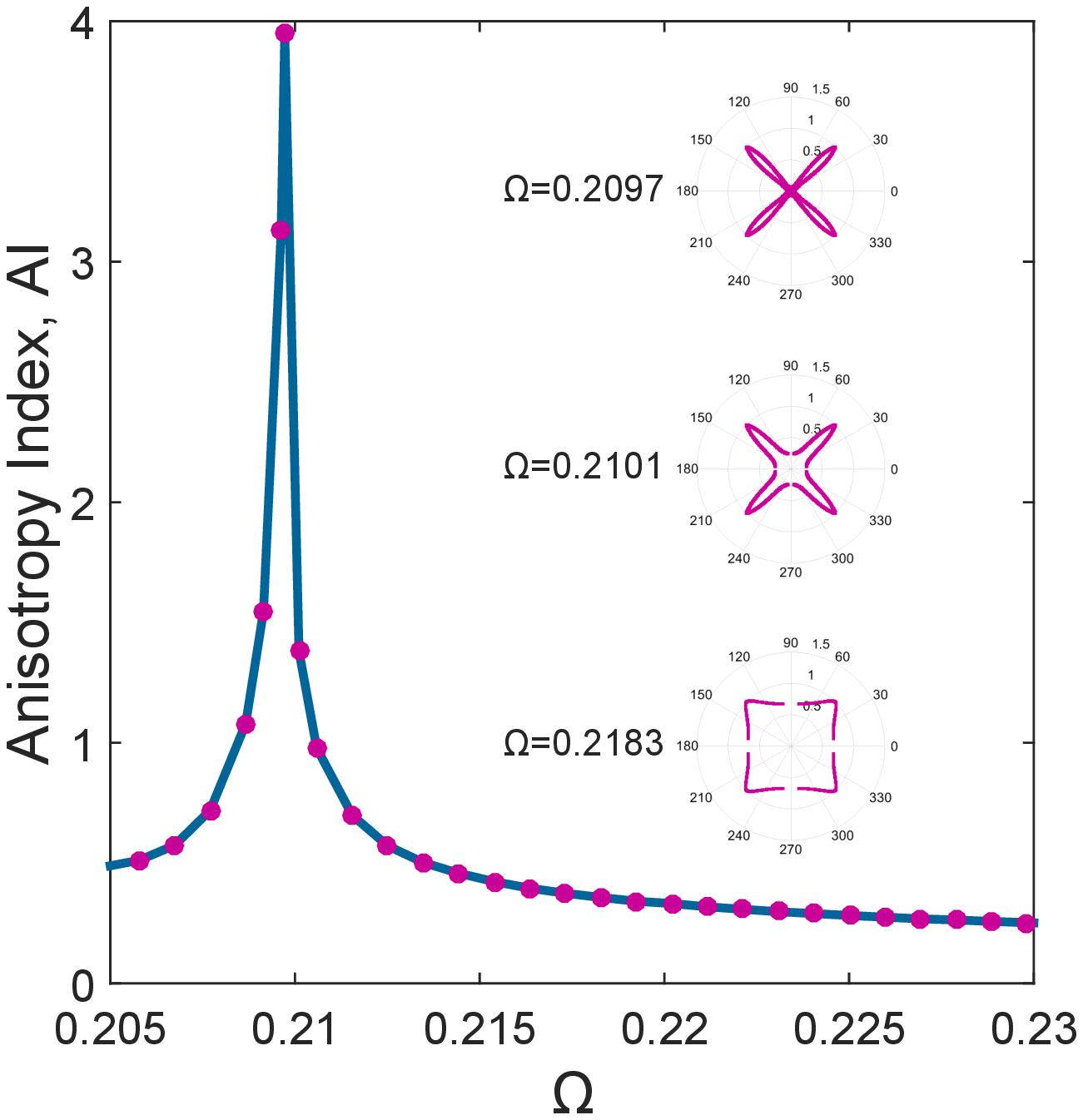}
                \caption{Anisotropy index.}
                \label{fig:AI_Conf2_Polar10905Hz10925Hz11350Hz}
        \end{subfigure}
        \quad \quad
        \begin{subfigure}[b]{0.5\textwidth}
                \includegraphics[width=\textwidth]{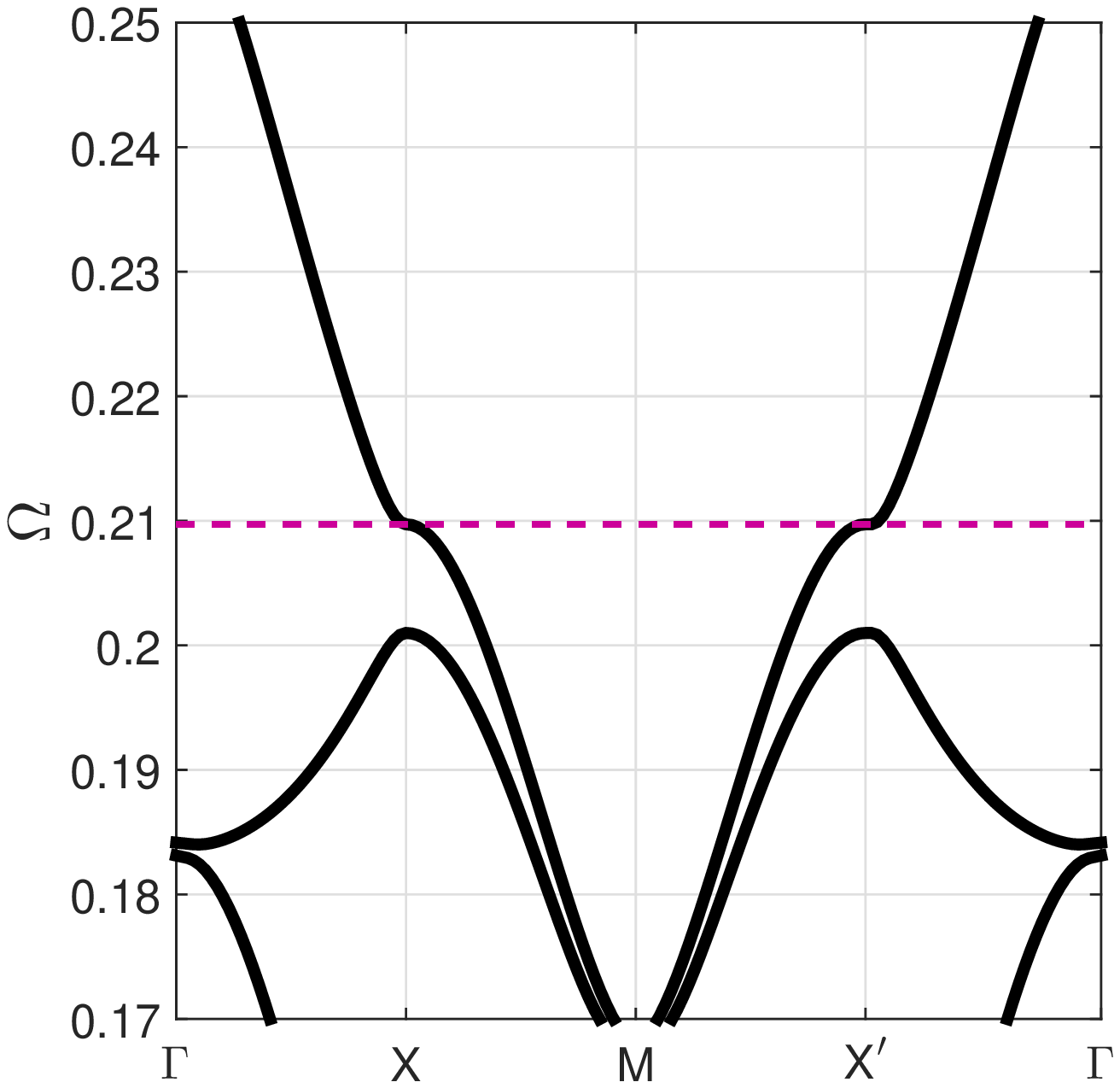}
                \caption{Band Diagram.}
                \label{fig:BD_UndConf2_Directionality}
        \end{subfigure}
                     
        \caption{Anisotropy index, polar group velocity plots for Configurations 2 (left) and detail of the band diagram (right). Geometric parameters: $\gamma=0.05$ and $\zeta=0.1$. Polar plots are normalized.}\label{fig:AI_BD_Conf2}
\end{figure}

From the definition of AI, it follows that $AI=0$ defines isotropic assemblies, while a structure having $AI\ne0$ is characterized by a certain degree of anisotropy. Frequencies corresponding to a directional behavior are investigated in the range $\Omega \in [0,0.35]$. The analysis of the band diagrams in Fig.~\ref{fig:BandDiagramConf1_6} and Fig.~\ref{fig:BandDiagramConf2_7} for configurations 1 and 2, respectively, reveals that there are several frequency ranges where no dispersion branches are found along the $X-M$ direction. Therefore, wave propagation is not allowed along the corresponding physical direction $\hat{\mathbf{i}}_2$. Some branches are also characterized by stationary points of inflection in $X$. The present analysis is limited to the stationary points of inflection of the dispersion branches that depart from the band gaps. For undulated configuration 1, these points are $\Omega=0.1594$ and $\Omega=0.3072$, while for undulated configuration 2 stationary points of inflection correspond to  $\Omega=0.2097$ and $\Omega=0.3079$. Results shown in Fig.~\ref{fig:AI_BD_Conf1}  and Fig.~\ref{fig:AI_BD_Conf2} are obtained by computing the group velocity for increasing values of the frequency parameter $\Omega$, then evaluating the $AI$ associated with the corresponding wave mode. In the considered frequency range, Fig.~\ref{fig:AI_Conf1_Polar15150Hz15900Hz15974Hz} and Fig.~\ref{fig:AI_Conf2_Polar10905Hz10925Hz11350Hz} show that for both undulated configurations strong anistropic behavior is signaled by a peak in the value of AI. This peak of $AI$ also falls within a frequency range where only one wave mode is present, as shown by the details of the band diagrams in Fig.~\ref{fig:BD_UndConf1_Directionality} and Fig.~\ref{fig:BD_UndConf2_Directionality}. More specifically, in the case of the undulated configuration 1, anisotropy has a maximum at $\Omega=0.3072$, while in the case of undulated configuration 2 the maximum is reached at $\Omega=0.2097$. Despite the fact that the two configurations display strong anisotropy at different frequencies, they are characterized by very close values of $AI$ and also by very similar directionality plots. Indeed, waves propagate in both cases strictly along the $\pm 45^\circ$ directions.


\subsubsection{Transient response to harmonic excitation}
\begin{figure}
        \centering
        \begin{subfigure}[b]{0.45\textwidth}
                \includegraphics[width=\textwidth]{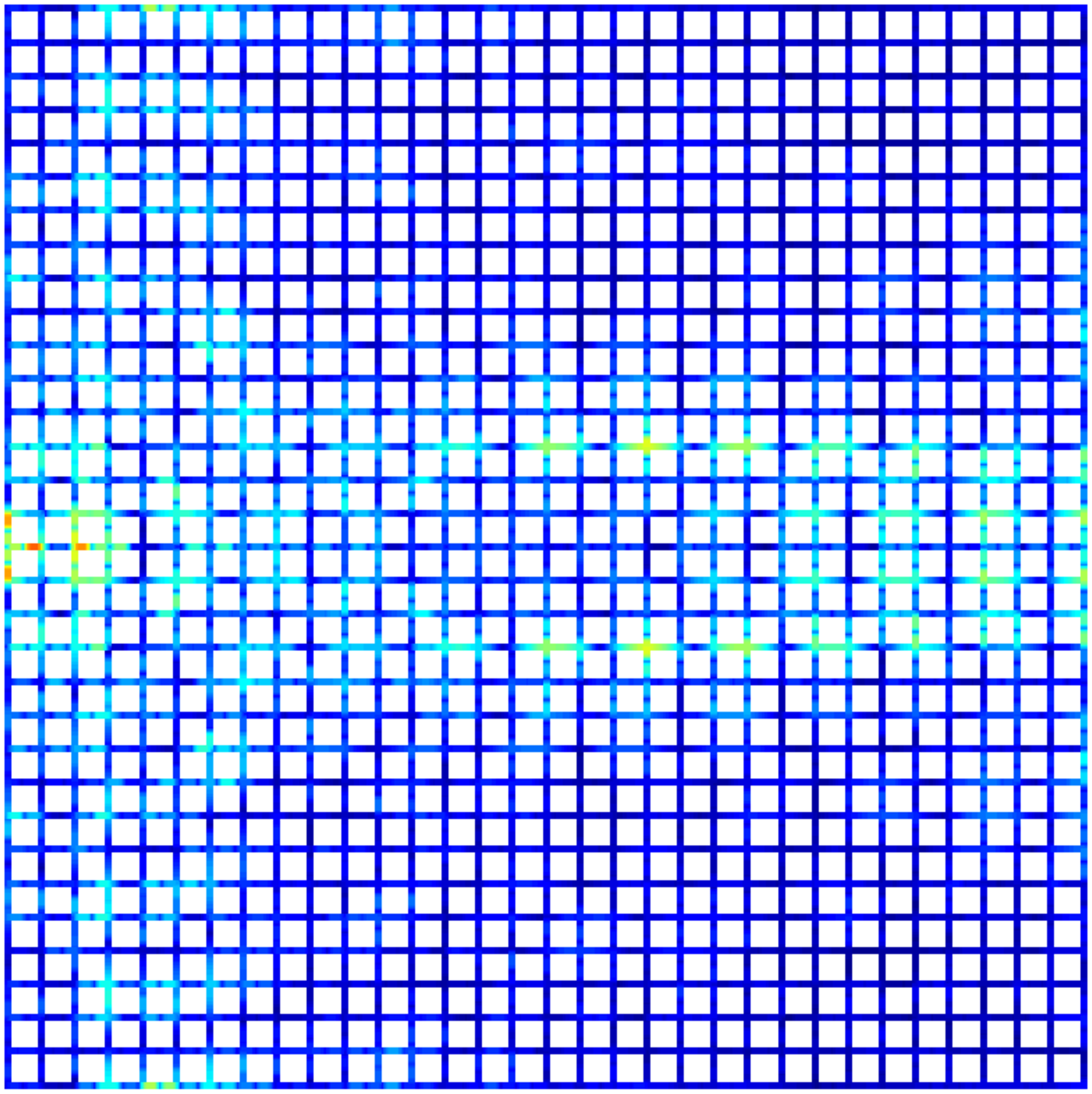}
                \caption{}
                \label{fig:TR_Straight_f15970Hz_U3_rot}
        \end{subfigure}\\
        \begin{subfigure}[b]{0.45\textwidth}
                \includegraphics[width=\textwidth]{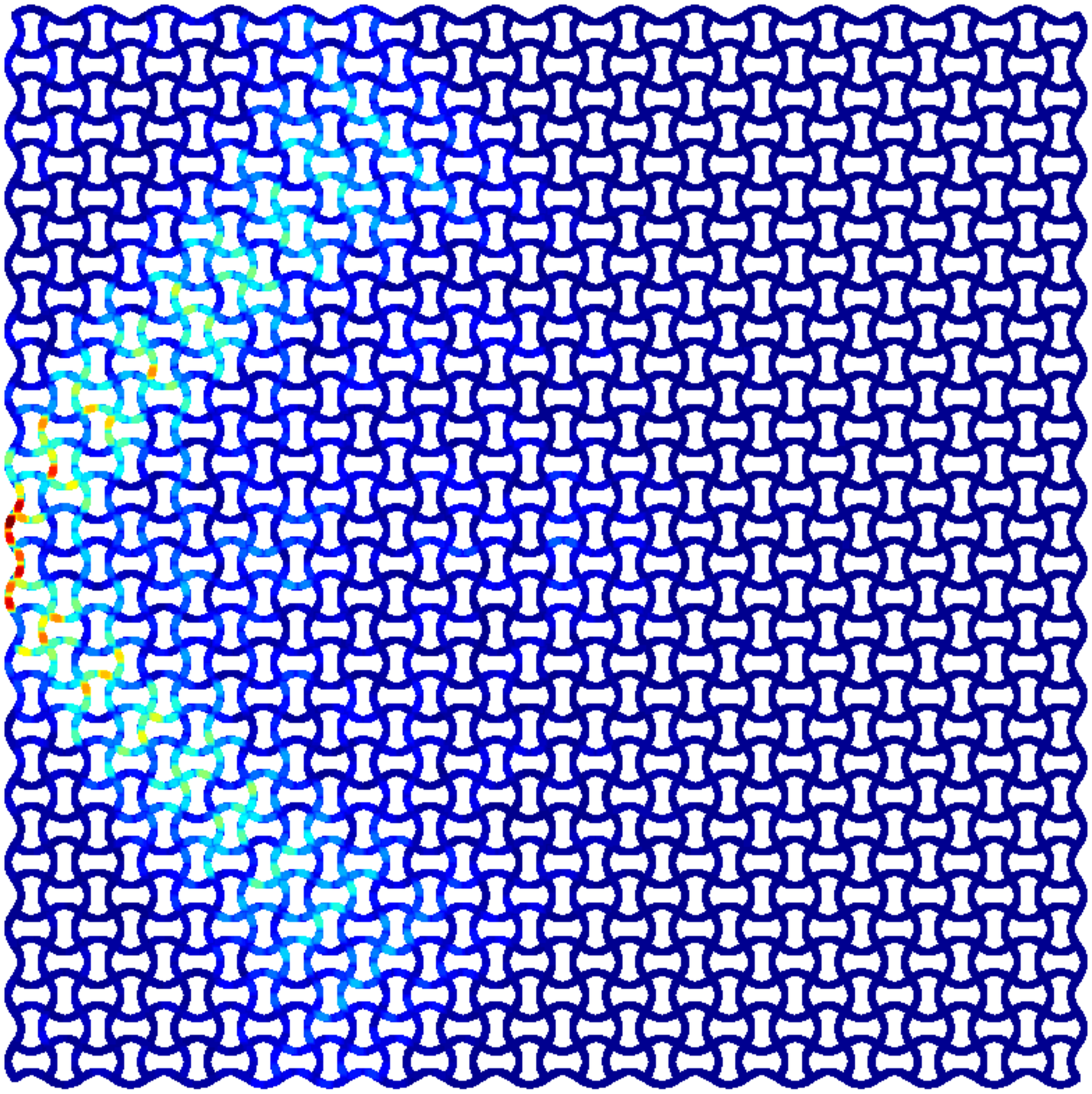}
                \caption{}
                \label{fig:TR_Conf1_f15970Hz_U3_rot}
        \end{subfigure}
            \begin{subfigure}[b]{0.45\textwidth}
                \includegraphics[width=\textwidth]{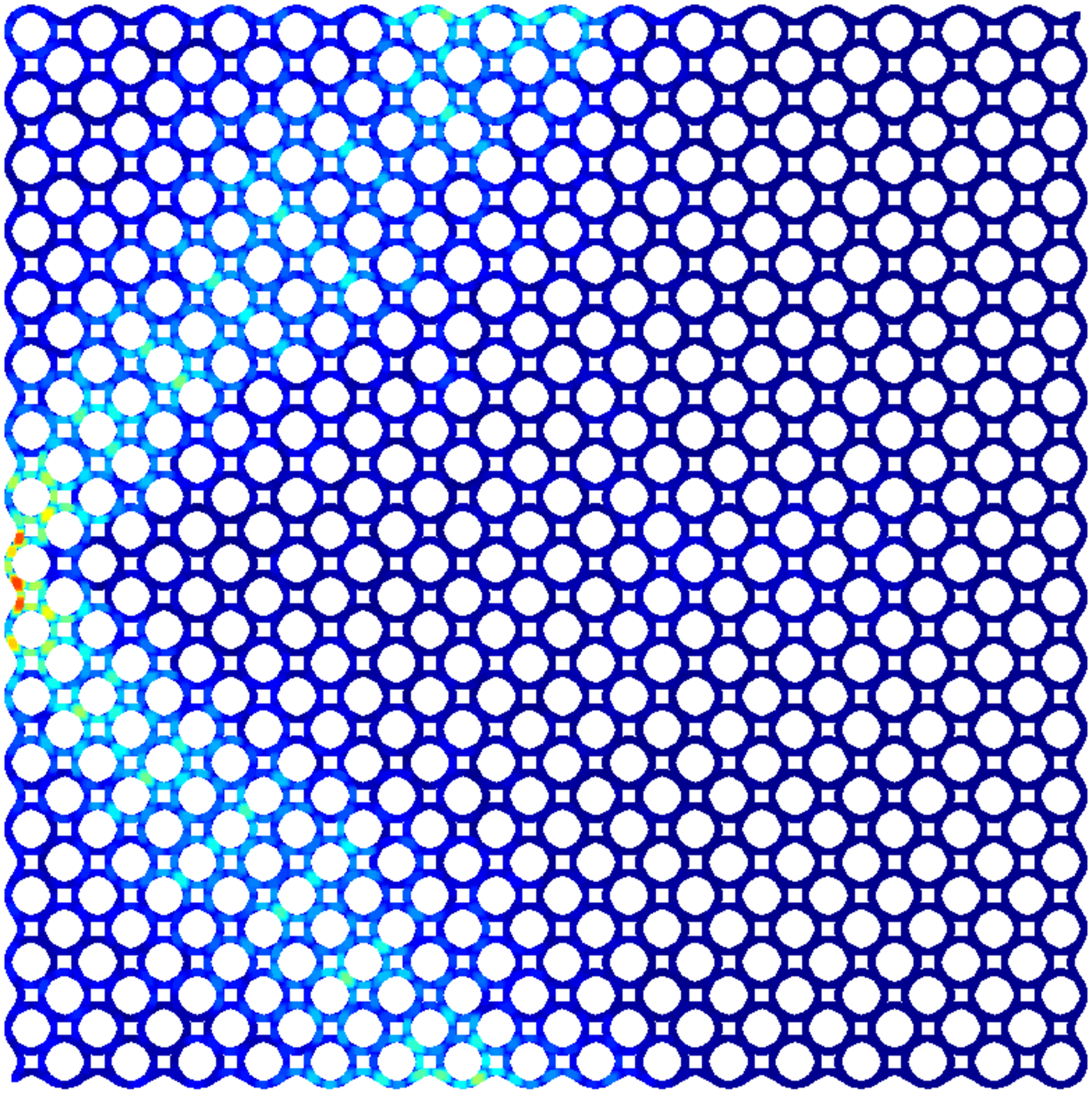}
                \caption{}
                \label{fig:TR_Conf2_f10900Hz_U3_rot}
        \end{subfigure}          
        \caption{Transient response to harmonic excitation for a square and undulated lattice of configuration 1 at $\Omega=0.3072$ (a,b), and for an undulated lattice of configuration 2 for $\Omega=0.2097$ (c) ($\zeta=0.1$ and $\gamma=0.05$).}\label{fig:TR_Conf1_f15970Hz}
\end{figure}

Finite 16x16 unit cell lattices are excited by imposing the local rotation of the lattice in a point placed in the mid span of one of the lattice's edges.  For both configuration 1 and 2, the geometric parameters are $\zeta=0.1$ and $\gamma=0.05$ . The excitation has the form of a sine wave with frequency corresponding to a selected value of $\Omega$, which in the case of configuration 1 is $\Omega=0.3072$, while for configuration 2 it is $\Omega=0.2097$. Equivalent square lattices are considered and their response is evaluated for the same excitation in order to provide a reference. Figure~\ref{fig:TR_Conf1_f15970Hz} show snapshots of the magnitude of the displacement field for the square and both the configuration 1 and 2 lattices. The displacement in each plot is normalized to the maximum valued at the considered time instant. Consistently with the group velocity predictions, undulated lattices allow energy to propagate only along the $\pm 45^\circ$ directions, while the energy propagates preferentially along the vertical and horizontal directions in the square lattice. Also, energy in undulated lattices travels at a lower speed compared to energy propagating in straight lattices.

\subsection{Graded undulated lattices}
Results developed in the previous sections can inform the design of structural lattices that are characterized by a spatial variation of the undulation parameter $\zeta$. The resulting graded lattices are non-periodic, and combine properties of straight and undulated  configurations by imposing selected variations $\zeta=\zeta(\mathbf{x})=\zeta(x_1,x_2)$. We show how graded lattices can be used to confine wave propagation within a certain region of a finite lattice. We consider two square graded lattices, one for each of the two considered configurations, having size $L_1=L_2=L$ and $\gamma=0.05$.

As a first example, the following undulation law is considered:
\[\zeta(x_1)=0.1\frac{x_1}{L}\]

\begin{figure}
\centering
                \includegraphics[width=0.8\textwidth]{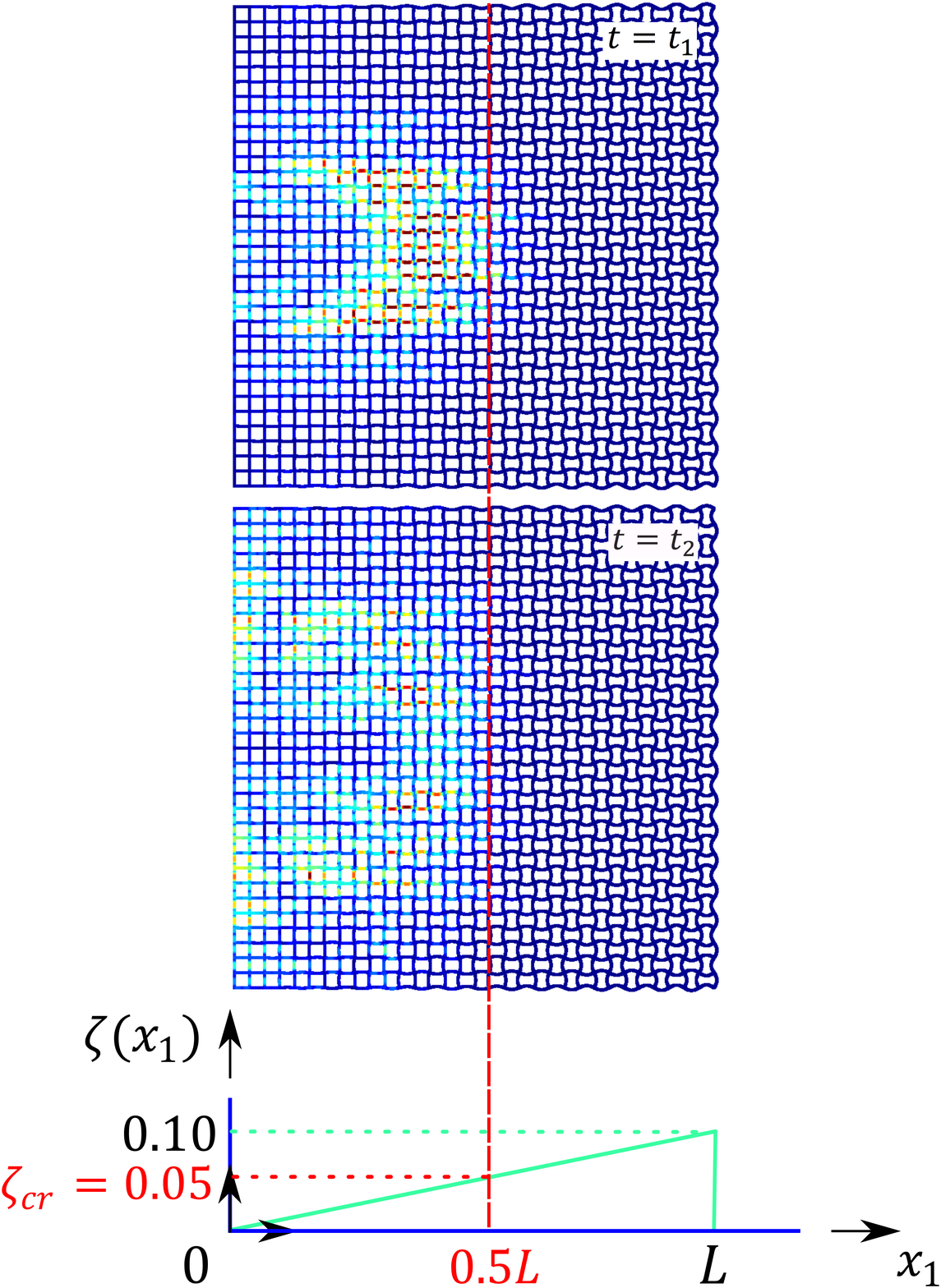}
                \label{fig:TR_StraightToConf2_f14039Hz_U2_Crop}
        	\caption{Finite straight-to-undulated lattice of configuration 1: two snapshots,  with $t_2>t_1$,  of transient response for excitation at $\Omega=0.23$ ($\gamma=0.05$ and undulation law $\zeta=0.1\frac{x_1}{L}$) show how wave propagation remains confined to the $0<x_1/L<0.5$ region.}\label{fig:TR_StraightToConf1_f11959Hz}
\end{figure}

\begin{figure}
\centering
                \includegraphics[width=0.8\textwidth]{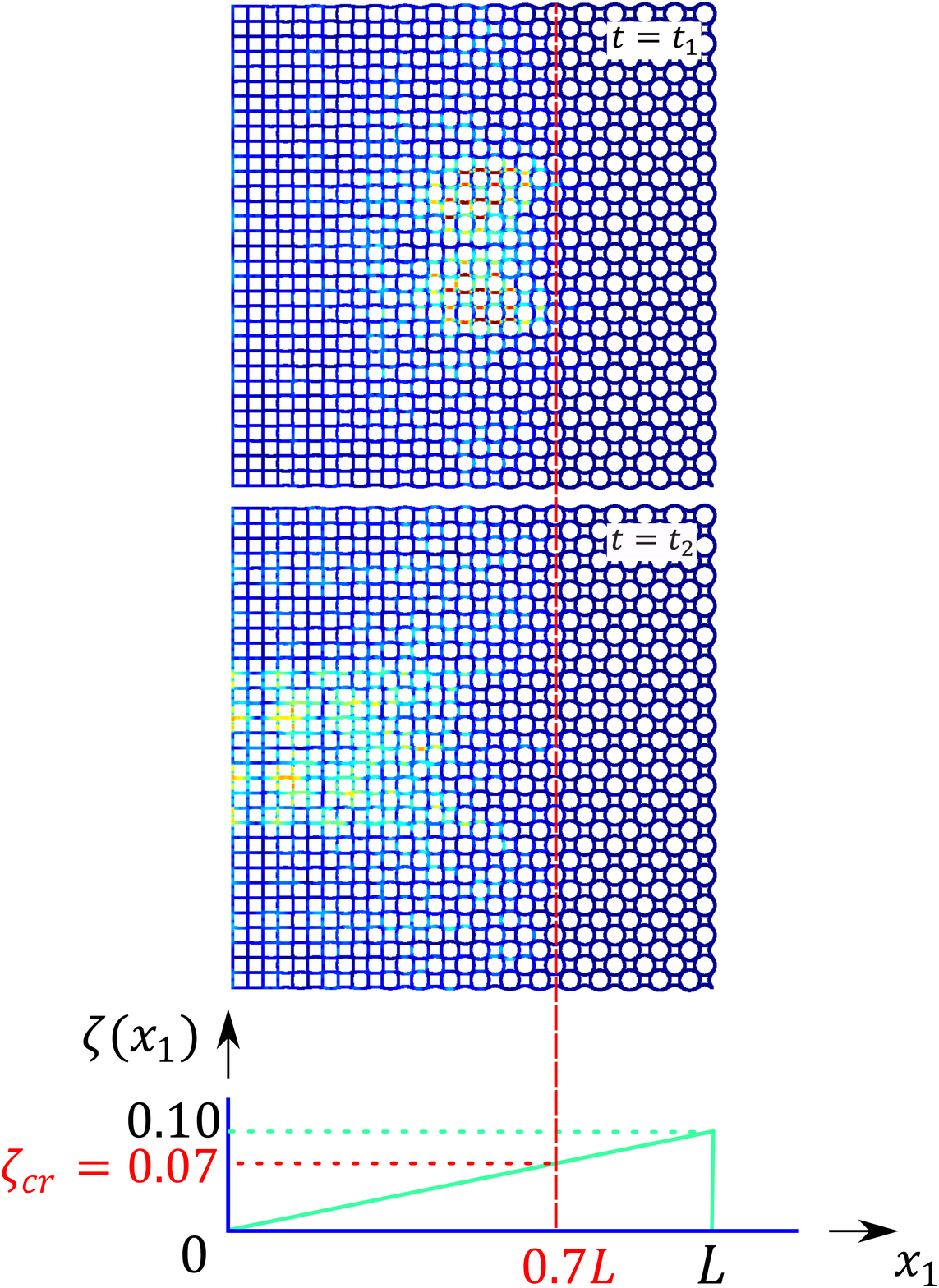}
                \label{fig:TR_StraightToConf2_f14039Hz_U3_Crop}
  	\caption{Finite straight-to-undulated lattice of configuration 2: two snapshots,  with $t_2>t_1$,  of transient response for excitation at $\Omega=0.27$ ($\gamma=0.05$ and undulation law $\zeta=0.1\frac{x_1}{L}$) show how wave propagation remains confined to the $0<x_1/L<0.7$ region.}\label{fig:TR_StraightToConf2_f11959Hz}
\end{figure}

The band gap map in Fig.~\ref{fig:BandGapMap_Conf1} shows that the critical undulation value $\zeta_c$, as illustrated in~\ref{BandDiagrams}, for configuration 1 is $\zeta_c\approx0.05$ at $\Omega=0.23$. Thus, an undulated lattice of configuration 1 having $\zeta_c>0.05$ does not support the propagation of waves propagating at this frequency. Transient response to narrow band excitation at $\Omega=0.23$ of a straight-to-undulated lattice is presented in Fig.~\ref{fig:TR_StraightToConf1_f11959Hz}. The simulation shows how the wave propagation is confined to the left part of the lattice, \emph{i.e.} $0<x_1/L<0.5$, being the right part of the lattice forbidden to the wave.  Similarly, the band gap map in Fig.~\ref{fig:BandGapMap_Conf2} shows that the critical undulation value $\zeta_c$ for the undulated lattice configuration 2 at $\Omega=0.27$ is approximately $\zeta_c=0.07$. For this reason, the undulated lattice with configuration 2 subject to a narrow band excitation at $\Omega=0.27$, shown in Fig.~\ref{fig:TR_StraightToConf2_f11959Hz}, displays a wave that initially propagates and then remains confined in $0<x_1/L<0.7$.

Graded lattices can also be used to modify the directionality of wave motion. This is illustrated in a second example where we consider two lattices for the two configurations, both characterized by $\gamma=0.05$. The undulation law is:
\begin{equation}
\zeta(x_1) = 
  \begin{cases} 
   0.1\frac{x_1}{L} & \text{if } 0<\frac{x}{L}<1 \\
   0.1(1- \frac{x_1}{2L})       & \text{if } 1<\frac{x}{L}<2
  \end{cases}
\end{equation}
Due to the imposed modulation of $\zeta=\zeta(x_1)$, the lattice goes from straight to undulated configuration, then from undulated to straight. Focus is placed on the low-frequency range, \emph{i.e.} $\Omega \in [0,0.05]$, where both configuration 1 and 2 lattices have an almost identical dynamic behavior, as one can inferred from the band structures of Fig.~\ref{fig:BandDiagram_wLTdisp}. Transient response to harmonic excitation is performed by imposing a displacement varying harmonically at $\Omega=0.040$ along the $\mathbf{i}_1$ direction, similarly to previous examples. As a result of the nature and direction of the excitation, a longitudinal wave is produced with a strong directional behavior along the $\hat{\mathbf{i}}_1$ direction. Then, for both configurations, the undulation is responsible of a bifurcation of the wave from occurring approximately at $\frac{x}{L}=0.4$ along $\pm 30^\circ$ directions. This behavior can be explained by looking at the group velocity plots of the lattices at $\Omega=0.040$ for different values of $\zeta$. We observe that increasing undulation is associated to the formation of lobes, signaling strong energy focusing.

\begin{figure}
        \centering
        \begin{subfigure}[b]{0.5\textwidth}
                \includegraphics[width=\textwidth]{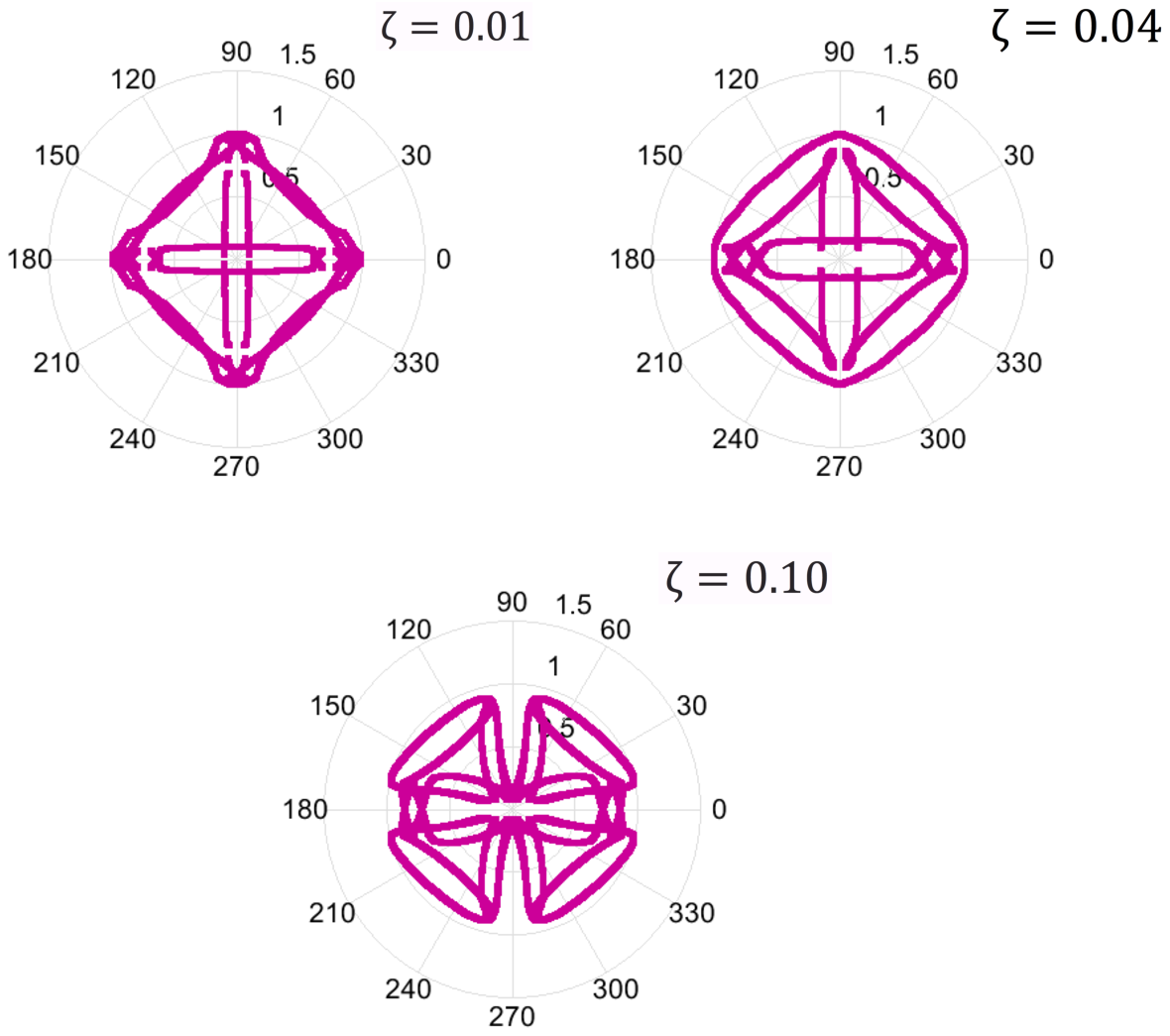}
                \caption{}
                \label{fig:Conf1_cg_SUS}
        \end{subfigure}\\
        
        \begin{subfigure}[b]{0.7\textwidth}
                \includegraphics[width=\textwidth]{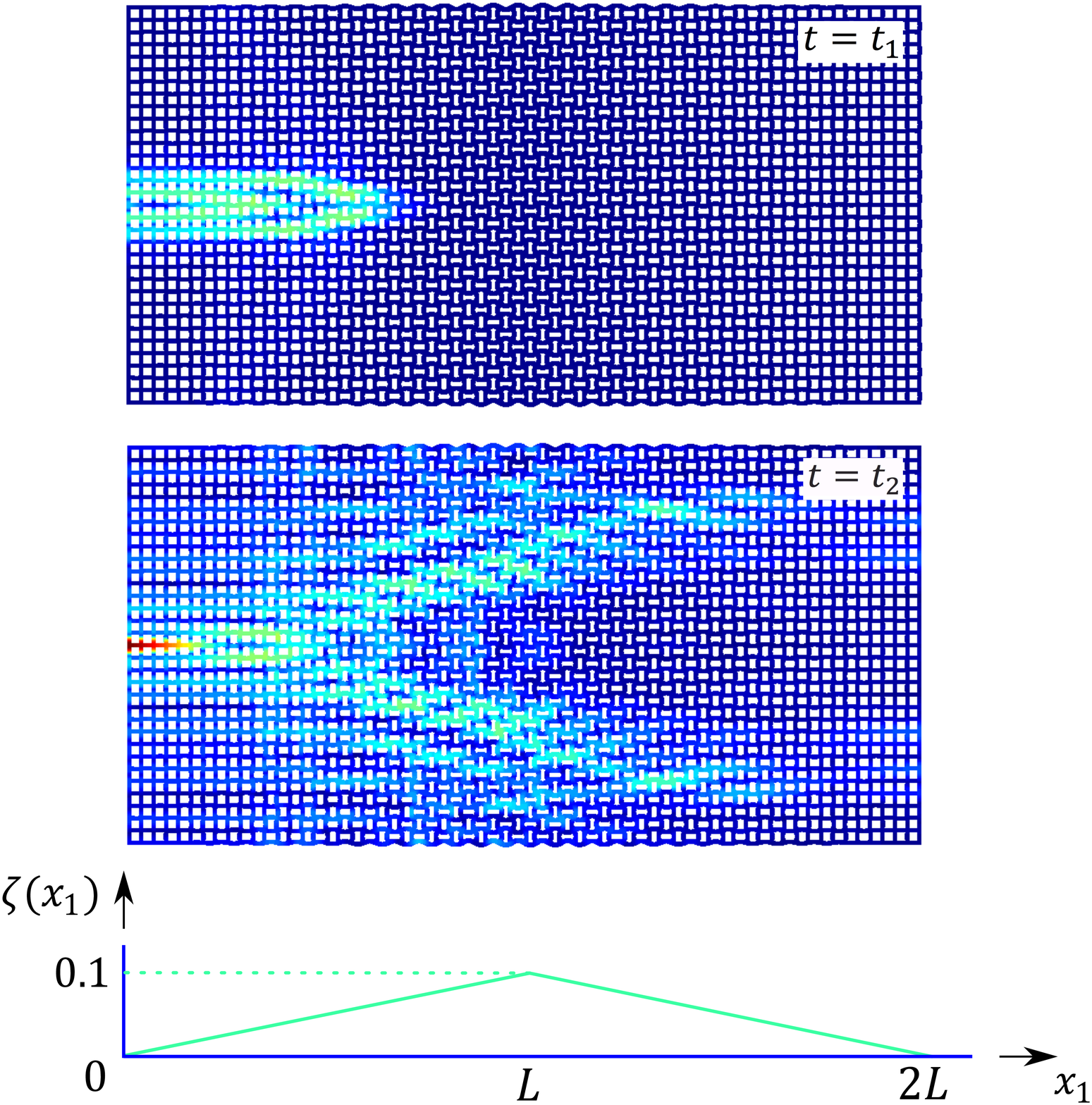}
                \caption{}
                \label{fig:SUS_Conf1_Alt_nocg}
        \end{subfigure}     
        \caption{Finite straight-undulated-straight lattice of configuration 1: (a) group velocity plots of periodic undulated lattices for different values of undulation amplitude $\zeta$; (b) two snapshots,  with $t_2>t_1$, of transient response for excitation at $\Omega=0.04$ ($\gamma=0.05$ and linear undulation law) show bifurcation of wave propagation starting at approximately $x_1/L=0.4$.}\label{fig:SUS_Conf1}
\end{figure}

\begin{figure}
        \centering
        \begin{subfigure}[b]{0.5\textwidth}
                \includegraphics[width=\textwidth]{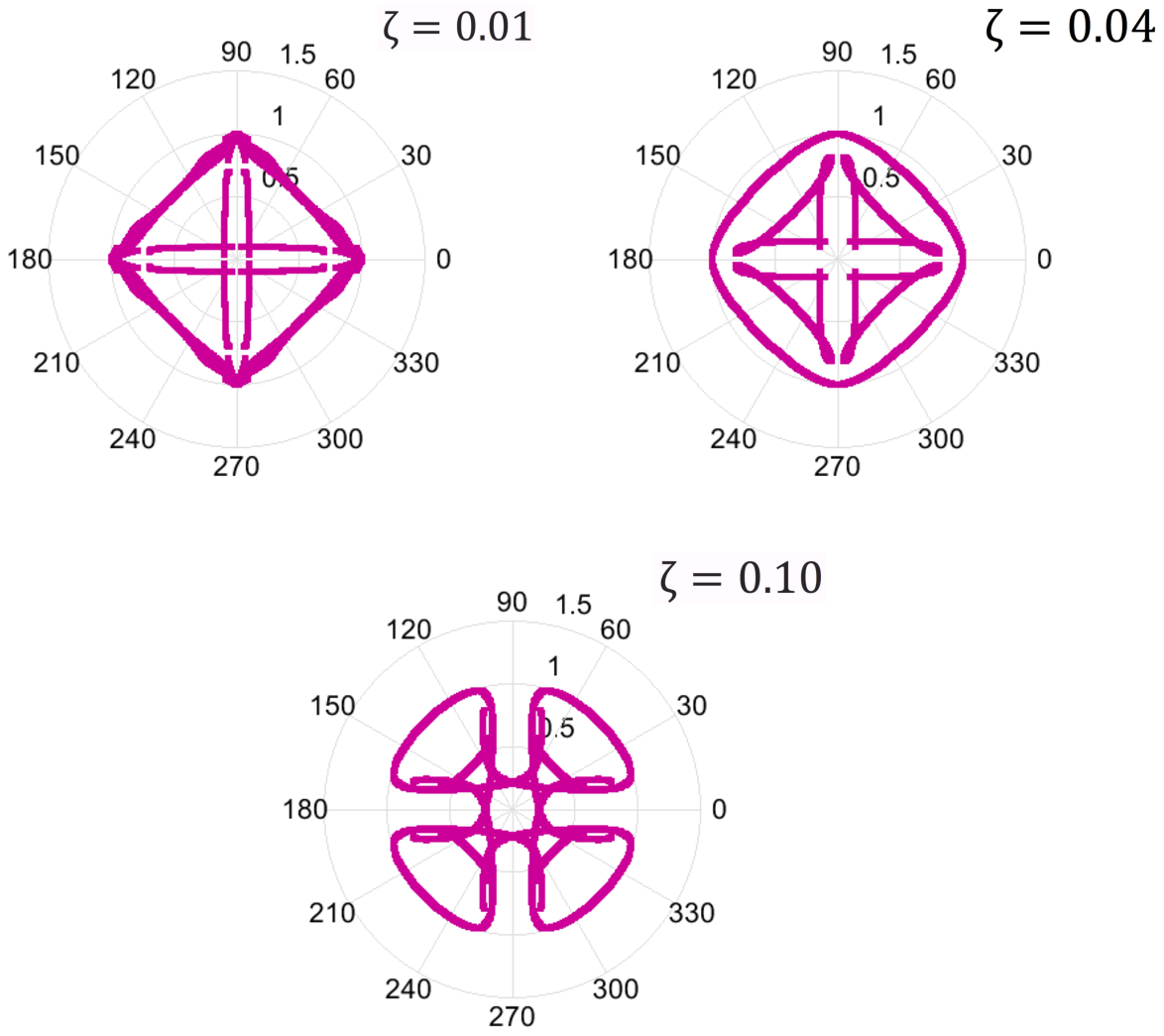}
                \caption{}
                \label{fig:Conf2_cg_SUS}
        \end{subfigure}\\
        
        \begin{subfigure}[b]{0.7\textwidth}
                \includegraphics[width=\textwidth]{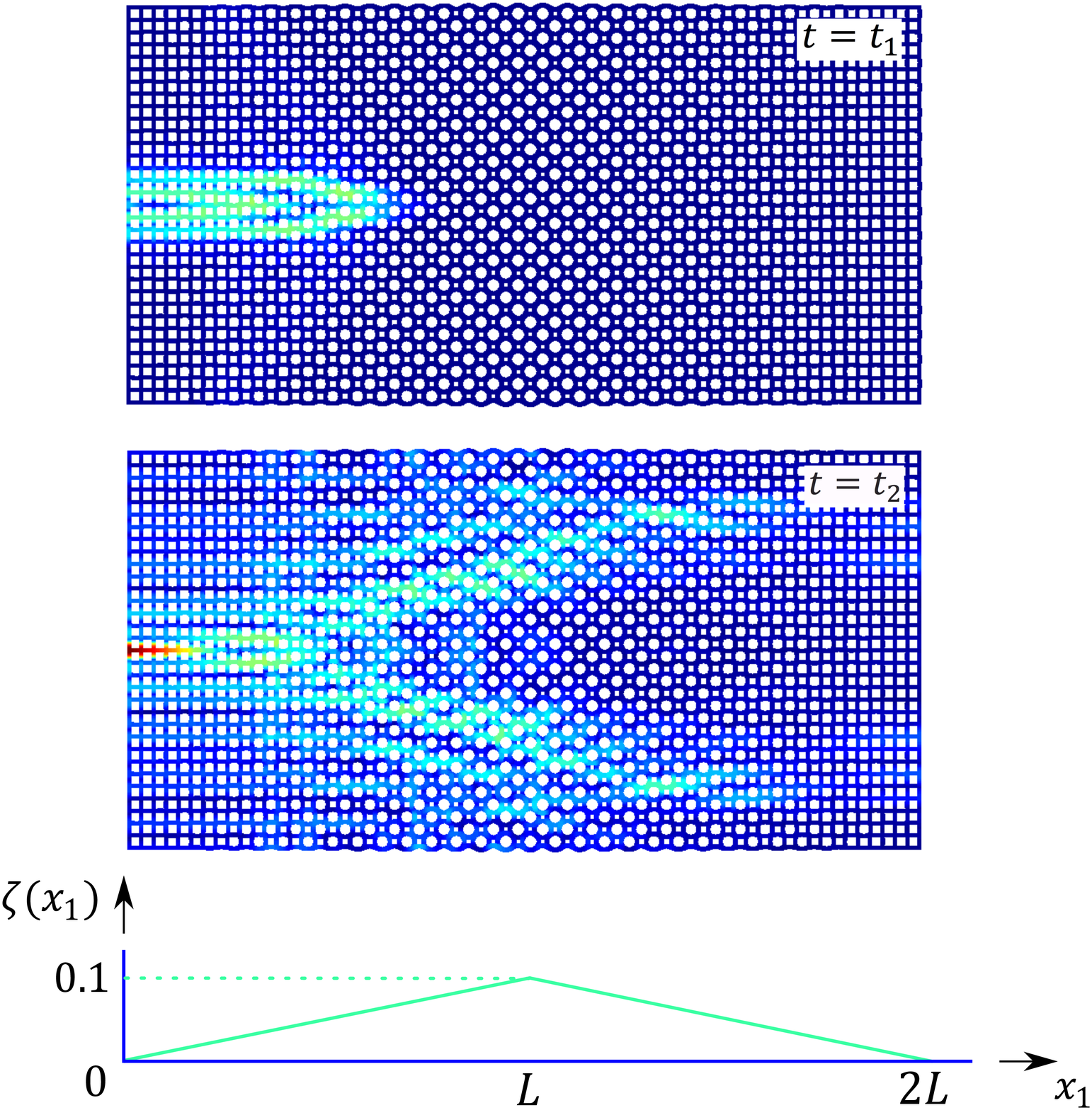}
                \caption{}
                \label{fig:SUS_Conf2_Alt_nocg}
        \end{subfigure}     
        \caption{Finite straight-undulated-straight lattice of configuration 2: (a) group velocity plots of periodic undulated lattices for different values of undulation amplitude $\zeta$; (b) two snapshots,  with $t_2>t_1$,  of transient response for excitation at $\Omega=0.04$ ($\gamma=0.05$ and linear undulation law) show bifurcation of wave propagation starting at approximately $x_1/L=0.4$.}\label{fig:SUS_Conf2}
\end{figure}

\section{Conclusions}
\label{sec:Conclusions}
This paper investigates wave propagation in square lattice structures with straight and undulated beams. Wave propagation properties are first studied numerically by performing Bloch analysis on the unit cell of the system. Band diagrams and group velocity plots have inform further analysis based on the simulation of wave propagation in finite lattices. The results of the analysis show that undulated lattices, contrarily to straight lattices, display band gaps in the considered frequency range. The influence of the geometric parameters on the amplitude of the band gaps is investigated and reported in the form of band gap maps, which unveil the drastically different dynamic behavior of the two undulated configurations in the high frequency range. When directionality of wave propagation is of interest, undulated lattices display single wave modes that tend to guide waves along specific directions. A quantitative representation of the directionality of selected wave-modes relies on the definition of an Anisotropy Index, which peaks at different frequency for the two configurations considered. Results have been numerically validated by computing the transient response to harmonic excitation of finite lattices. Finally, graded lattice configuration illustrate the possibility to conveniently modulate the structure's undulation to define areas of the lattice where wave propagation, and the energy associated to it, can be confined. Similarly, modulation of the undulation can be used together with the knowledge of the group velocity plots to affect the directionality properties of the structure and induce, for example, wavefront splitting. The discussed results show how graded structural lattices have a great potential for wave propagation control, suggesting the possibility of broadband filtering by combining modulation of undulation and thickness.

\section{Acknowledgments}
\label{sec:Acknowledgments}
\emph{This work is conducted under the support of the Army Research Office, under grant No. W911NF1210460 monitored by Dr. David Stepp, whose support is greatly appreciated.}

\clearpage

\newpage

\bibliographystyle{unsrt}
\clearpage
\bibliography{PaperLattices_ArXiv}

\end{document}